\documentstyle[useAMS,graphicx,longtable]{mn2e}

\newcommand{\hMpc}{{\ifmmode{h^{-1}{\rm Mpc}}\else{$h^{-1}$Mpc}\fi}}
\newcommand{\hkpc}{{\ifmmode{h^{-1}{\rm kpc}}\else{$h^{-1}$kpc}\fi}}

\def\approxlt{\mathrel{\spose{\lower 3pt\hbox{$\sim$}}
        \raise 2.0pt\hbox{$<$}}}
\def\approxgt{\mathrel{\spose{\lower 3pt\hbox{$\sim$}}
        \raise 2.0pt\hbox{$>$}}}
\def\approxpropto{\mathrel{\spose{\lower 3pt\hbox{$\sim$}}
        \raise 2.0pt\hbox{$\propto$}}}

\title{Eclipsing binaries in ASAS catalog}
\author[B. Paczy\'nski, D. Szczygiel, B. Pilecki, and G. Pojma\'nski]
{B. Paczy{\'n}ski$^{1}$\thanks{e-mail: bp@astro.princeton.edu}
and D. Szczygiel$^{2}$, B. Pilecki$^{2}$, G. Pojma\'nski$^{2}$
\thanks{e-mail:  dszczyg, pilecki, gp@astrouw.edu.pl}\\
$^{1}$Princeton University Observatory, Peyton Hall, Princeton, NJ 08544, USA\\
$^{2}$Warsaw University Observatory, Al. Ujazdowskie 4, PL-00-478, Poland}

\date{Accepted --.
      Received -- ;
      in original form --}

\pubyear{2005}

\begin{document}

\maketitle

\label{firstpage}

\begin{abstract}
ASAS is a long term project to monitor bright variable stars
over the whole sky.  It has discovered 50,122 variables brighter than
${\rm V < 14 }$ mag south of declination ${\rm + 28^{\circ}}$, and among them
11,099 eclipsing binaries.  We present a preliminary analysis
of 5,384 contact, 2,957 semi-detached, and 2,758 detached systems.
The statistics of the distribution provides a qualitative confirmation
of decades old idea of Flannery and Lucy that W UMa type binaries
evolve through a series of relaxation oscillations:
ASAS finds comparable number of contact and semidetached systems.  

The most surprising
result is a very small number of detached eclipsing binaries with
periods ${\rm P < 1 }$ day, the systems believed to be the progenitors
of W UMa stars.  As many (perhaps all) contact binaries have companions,
there is a possibility that some were formed in a Kozai cycle, as
suggested by Eggleton and his associates. 
\end{abstract}


\section{Introduction to ASAS}  
\label{sect:intro1}

ASAS - All Sky Automated Survey, is a long term project dedicated
to detection and monitoring variability of bright stars.  This
paper presents the results of several years of observations
done at the Las Campanas Observatory with a single instrument:
a telescope with the aperture of 7 cm, the focal length of 20 cm,
done through a standard V-band filter and a
${\rm 2K \times 2K }$ CCD camera with ${\rm 15 \mu }$ pixels
from Apogee (Pojma\'nski 1997, 1998,
2000, 2002, 2003, Pojma\'nski and Maciejewski 2004, 2005, 
Pojma\'nski, Pilecki and Szczygiel 2005).  More information 
about ASAS is provided on the WWW:

\centerline{ http://www.astrouw.edu.pl/\~{}gp/asas/asas.html }
\centerline{ http://archive.princeton.edu/\~{}asas/ }

\begin{figure*}
\vspace{-0.9\linewidth}
\includegraphics[width=\linewidth]{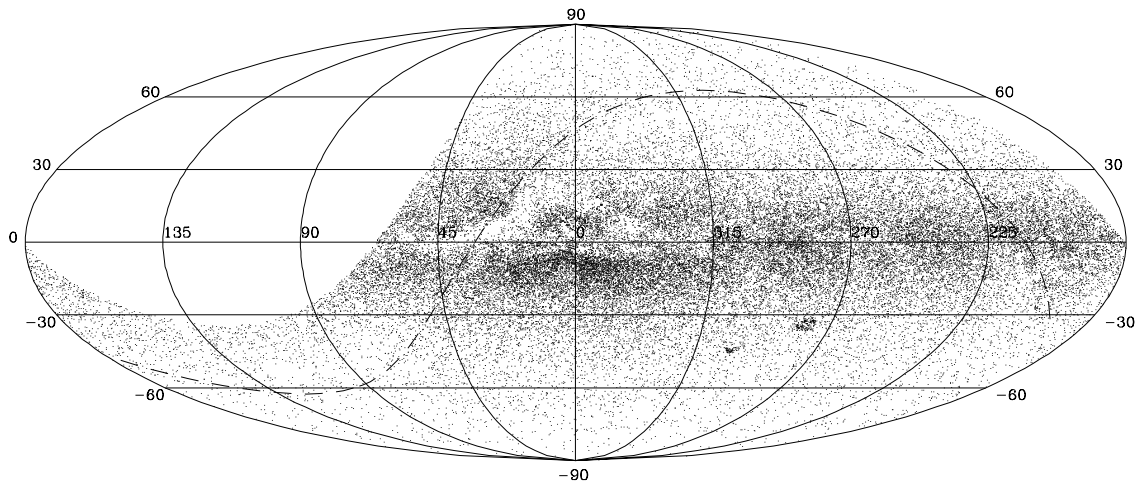}
\caption{
The distribution of 50,122 ASAS variables in the sky in Galactic
coordinates.  The Milky Way is clearly seen, as well as the patches
of interstellar extinction.  The equator is shown with a dashed line.
The distribution is limited to declination ${\rm < +28^{\circ}}$.
}
\end{figure*} 

The variable stars were discovered quasi - uniformly for declination
${\rm < + 28^{\circ} }$, covering almost 3/4 of the full sky.
Fig. 1 shows the distribution of ASAS variables in the sky in the
Galactic coordinates.  The Milky Way is clearly visible, together
with the dust lanes.  The distribution of ASAS stars as a function of 
V-mag is shown on Fig. 2 for all stars (17,000,000), all
variable stars (50,122), and all eclipsing binaries,
the latter divided into Eclipsing Contact binaries (EC, 5,384),
Eclipsing Semi Detached binaries (ESD, 2,957), and Eclipsing Detached 
binaries (ED, 2758).  
The statistics for stars with ${\rm 8 < V < 12 }$ mag appears to be
approximately complete, but the efficiency falls rapidly within
the range of ${\rm 12 < V < 14 }$ mag as the detection limit
is approached.  Also, the statistics deteriorates for ${\rm 8 < V }$
because of saturation  effects.

\begin{figure}
\includegraphics[width=\linewidth]{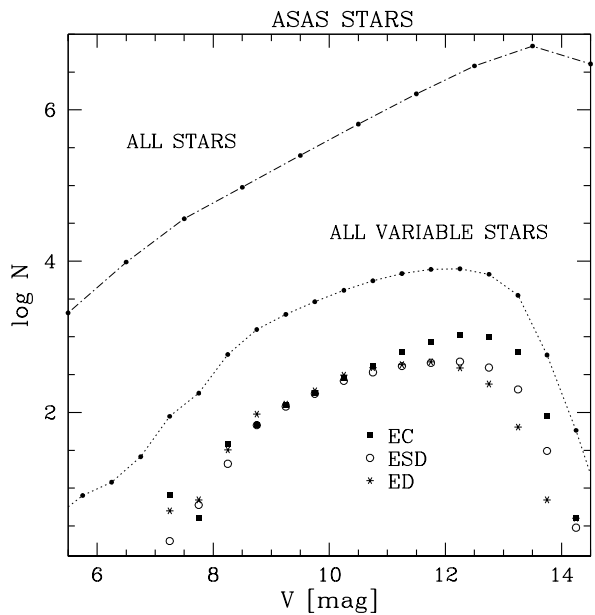}
\caption{
The distribution of ASAS stars as a function of V-band magnitude.
The total number of stars is about 17,000,000.  The total number
of variable stars is 50,122.  The total number of eclipsing
binaries is 5,384 for contact systems (EC), 2,957 for semi-detached
systems (ESD), and 2,758 for detached systems (ED).  Notice, that
the efficiency of discovering variable stars declines for ${\rm V > 12 }$ 
because the detection limit is approached, and for ${\rm V < 8 }$ because 
of saturation effects.
}
\end{figure}

All stars were observed for about 5 years, with a small subset for 8 years.
Typical number of photometric measurements was several hundred.  The
distribution of this number is shown in Fig. 3. 
The total number of photometric measurements was 2,916,000.
As ASAS continues its operation the number of measurements will increase, 
approximately 100 V-band photometric measurements per year per variable.
We intend to continue the project indefinitely, with
some upgrades.  While only V-band results are reported here,
the I-band photometry was accumulated, and several years of data are
already stored on a RAID-5 disk system.  However, it will take 
another year to process I-band data, well over a Tera-byte.  There
are more stars detectable in the I-band, so we expect that the number of 
variables will more than double.  We are also planning an expansion of 
ASAS to the northern hemisphere, to fully cover the whole sky.  

\begin{figure}
\includegraphics[width=\linewidth]{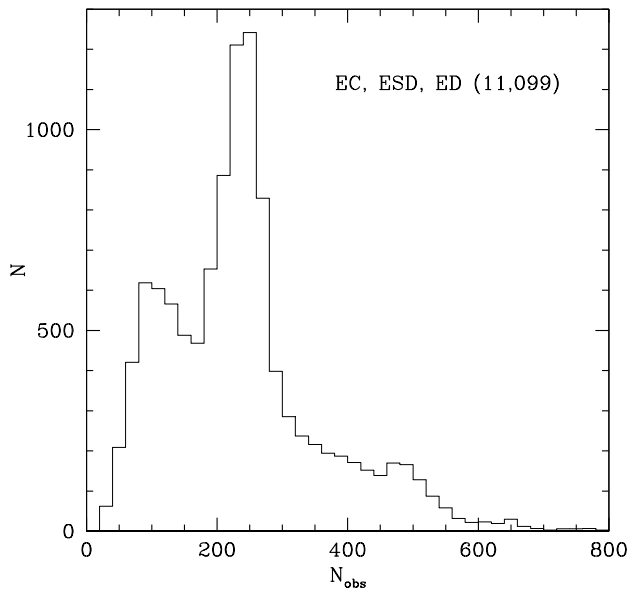}
\caption{
The histogram of the number of photometric measurements obtained
during 5 years of ASAS life.  A small subset of data extends
back to 8 years.
}
\end{figure}

The results presented in this paper are not final in any sense.  
Please note, that all data are public domain.  A different 
classification of binaries can be readily done by whoever feels
like verifying and/or correcting our presentation. 

This is an observational paper, but section 2 gives
a short introduction to the main ideas about the structure and evolution
of contact binaries.  Section 3 gives the information about ASAS data,
the classification scheme, and some examples of a diversity of
light curves.  Section 4 provides simple statistics of ASAS binaries.
Finally, in section 5 we make a somewhat speculative discussion based on this
observational paper.

\section{Introduction to contact binaries}
\label{sect:intro2}

Contact binaries, also known as W UMa stars, are in a physical
contact, with continuously changing brightness because of large tidal 
distortion of the two components.

The first theoretical milestone in the understanding of contact
binaries was due to Lucy (1968a,b), who proposed that the two components
share common envelope with the same entropy, thereby 
making the effective temperature almost constant over the surface of the
two stars.  As contact binaries have the mass ratio distinctly
different from one, most nuclear energy is generated in the
more massive component and it is redistributed around the
whole surface through a moderately thick convective envelope.

The second theoretical milestone was the recognition of the
consequences of the fact that the mass - radius relation for Zero
Age Main Sequence (ZAMS) stars is much steeper than for the
two Roche lobes.  There can be no stable equilibrium between the
two stars with a common envelope.
The system evolves through a sequence of relaxation
oscillations, with the mass flowing from star A to B,
next from B to A, etc. (Flannery 1976, Lucy 1976, Robertson
and Eggleton 1977).
The cycle repeats on a thermal (Kelvin-Helmholtz) time scale.
According to thermal relaxation model the binary oscillates
between thermal contact, with the two eclipses of almost
equal depth, and a semidetached phase in which one eclipse
is much deeper than another.

Hazlehurst (1970) suggested that nuclear evolution of the
primary component of a contact binary affects its structure.
Stepie\'n (2003, 2005)
suggested that the currently more massive primary was
originally the less massive of the two.  The nuclear burning formed 
a small helium core, the star expanded and transferred mass
to the original secondary.  In analogy with Algol systems the currently
more massive component is the less evolved, while the present
secondary has a small helium core, and it is more advanced in
its nuclear evolution.

It is interesting that there is a controversy about thermal relaxation
oscillations in W UMa systems.  Some authors claim there are no such
oscillations (Webbink 2003), while others claim that such
oscillations exist (Quin 2003, Yakut and Eggleton 2005,
and references therein).  ASAS statistics resolves this
controversy on purely observational grounds.

\section{ASAS data}
\label{sect:statistics}

Close binaries with a deep common envelope are in thermal contact
and they have eclipses of almost equal depth.
If the contact is shallow, or if there
is no physical contact, then the effective
temperatures of the two stars are different, 
and the two eclipses have different depth. 

Theoretical models of relaxation oscillations indicate that
the radii of the two components change relatively little 
throughout the cycle
(Flannery 1976, Lucy 1976, Robertson and Eggleton 1977,  Yakut
and Eggleton 2005, and references therein).
With the geometry of the two stars almost unchanged, tidal
distortions due to geometry remain almost the same,
and the most profound difference in the light curve
is the relative depth of the two eclipses. 

\begin{figure}
\includegraphics[width=\linewidth]{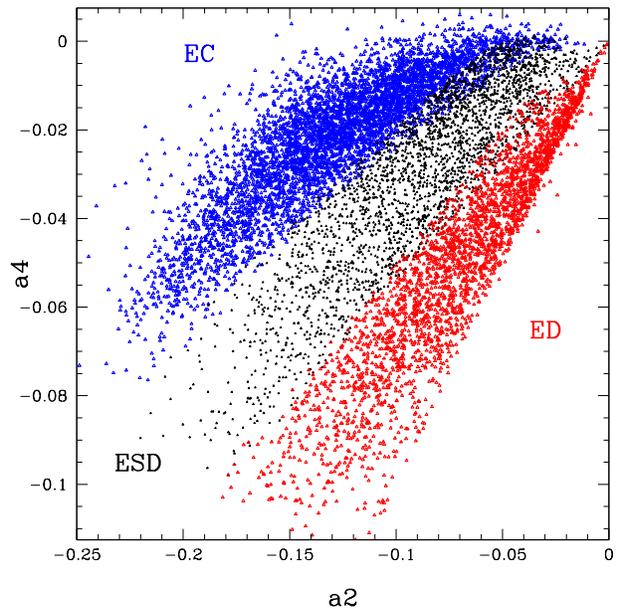}
\caption{
Classification of eclipsing binaries in the Fourier coefficients
plane ${\rm a_2 - a_4 }$.  The three symbols refer to
contact (EC), semidetached (ESD), and detached (ED)
binaries, following Pojma\'nski (2002).  Note: when the amplitude
of variability is very small the classification is very
uncertain, as in the upper right hand corner.
}
\end{figure}

For the purpose
of this paper the classification of eclipsing binaries was done
by decomposing their light curves into Fourier coefficients:
${\rm a_1, a_2, a_3, a_4 }$, following Ruci\'nski (1997a,b, 1998, 
2002) and Pojma\'nski (2002).  The
classification is shown in Fig. 4 in the ${\rm a_2 - a_4 }$ plane,
with the EC, ESD, and ED binaries shown with different symbols:
ESD stars have smaller marks than EC and ED. 
Some examples of bright binaries of EC, ESD, and ED type are
shown in Fig. 5.  The parameters of these 15 binaries are listed
in Table 1. The type of an eclipsing binary, EC, ESD, or ED, is
indicated at the upper right corner of the light curve. The complete
version of Table 1, for all eclipsing binaries used in this paper,
and with all eight Fourier parameters, is in the file {\tt Fourier.E}
on the web page:

\centerline{ http://www.astrouw.edu.pl/\~{}gp/asas/asas\_paper\_data.html }
\centerline{ http://archive.princeton.edu/\~{}asas/asas\_paper\_data.html }

\noindent
where all the data on which this paper is based are available electronically.
These include all photomertic measurements in compressed files: ec.tgz
(49MB), esd.tgz (27MB), ed.tgz (27MB). 

\begin{figure*}
\vspace{-0.12\linewidth}
\begin{tabular}{ccc}
\vspace{-0.12\linewidth}
  \resizebox{0.27\linewidth}{!}{\includegraphics*{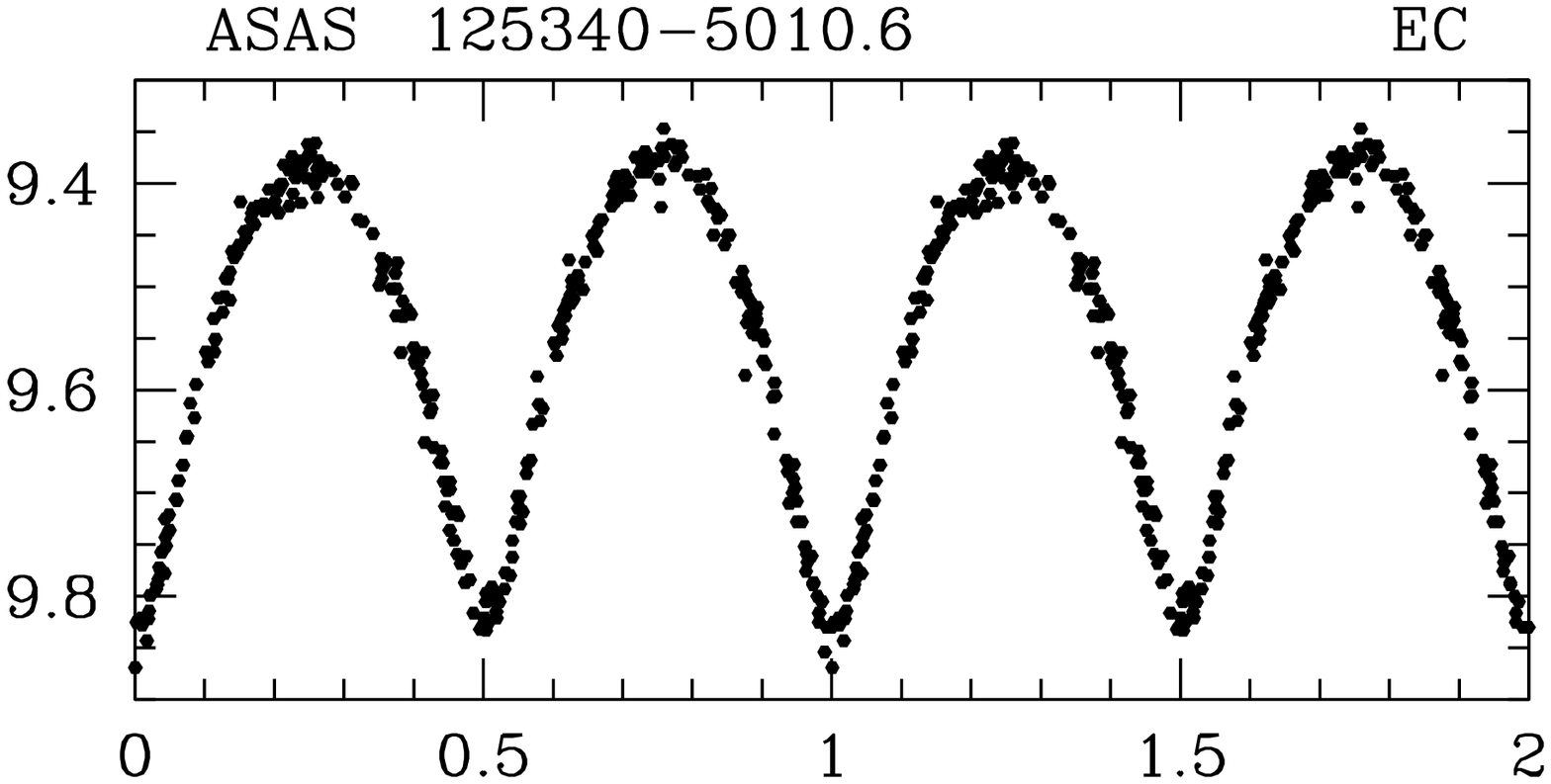}}&
  \resizebox{0.27\linewidth}{!}{\includegraphics*{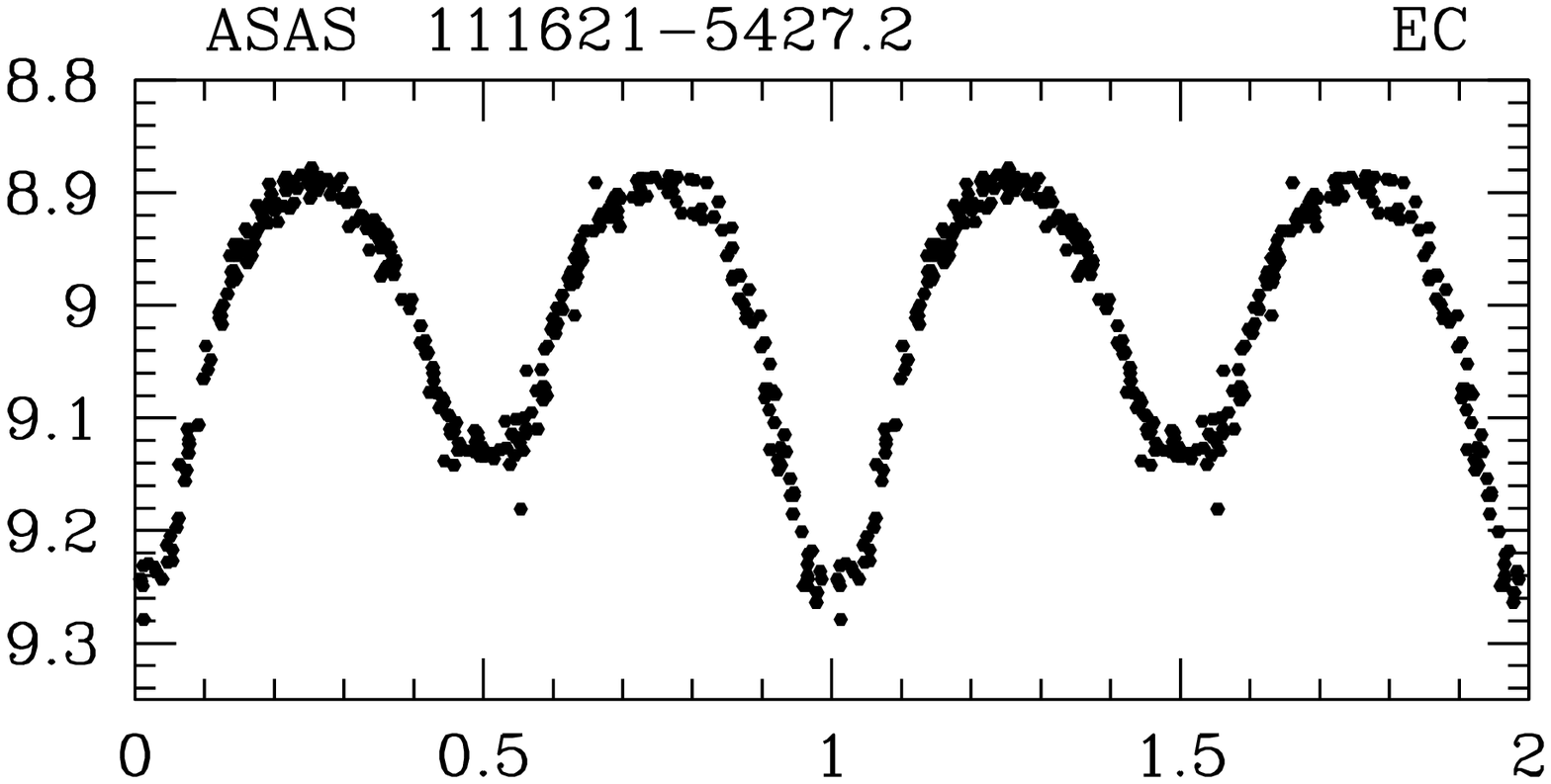}}&
  \resizebox{0.27\linewidth}{!}{\includegraphics*{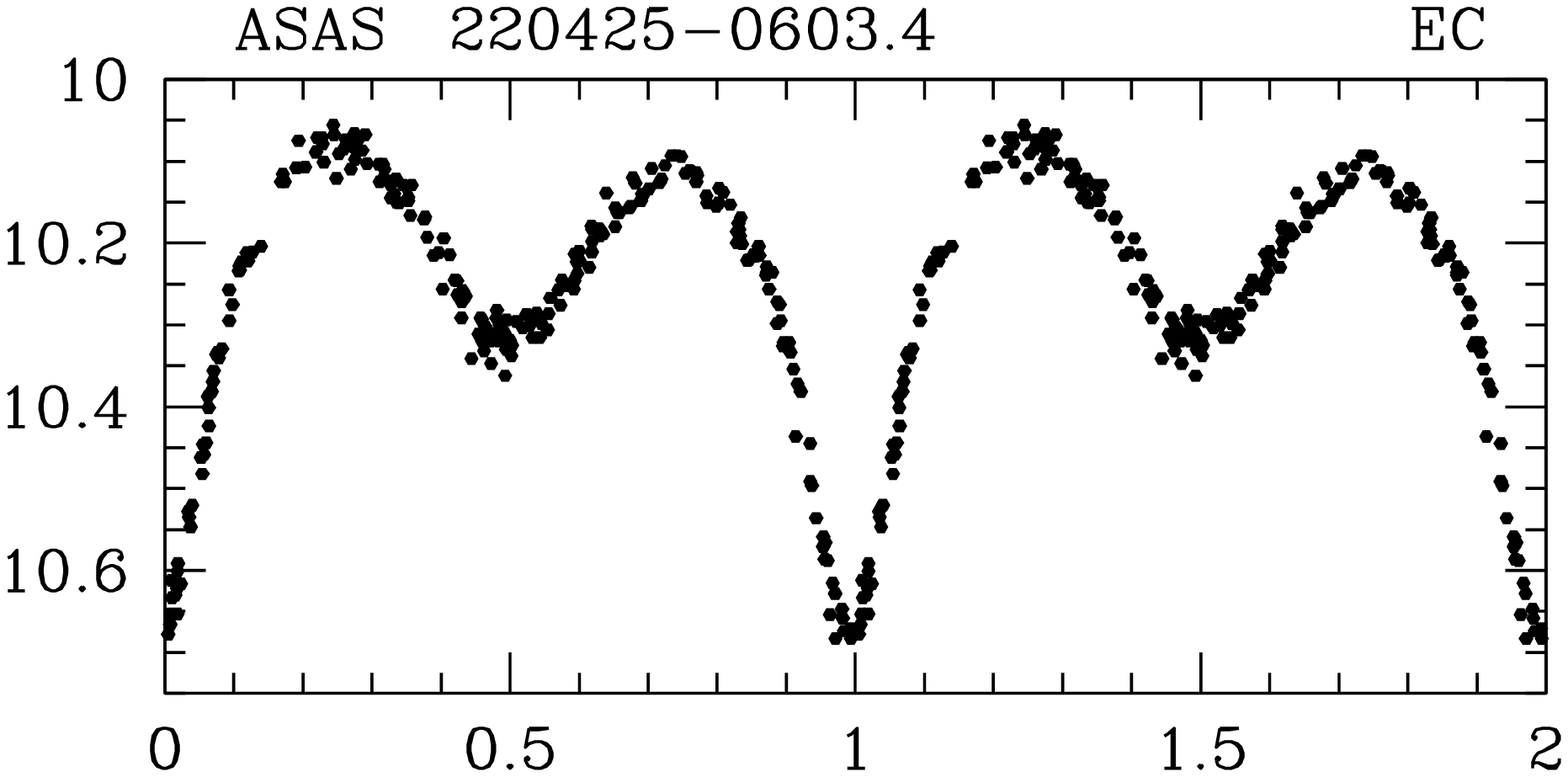}}\\
\vspace{-0.12\linewidth}
  \resizebox{0.27\linewidth}{!}{\includegraphics*{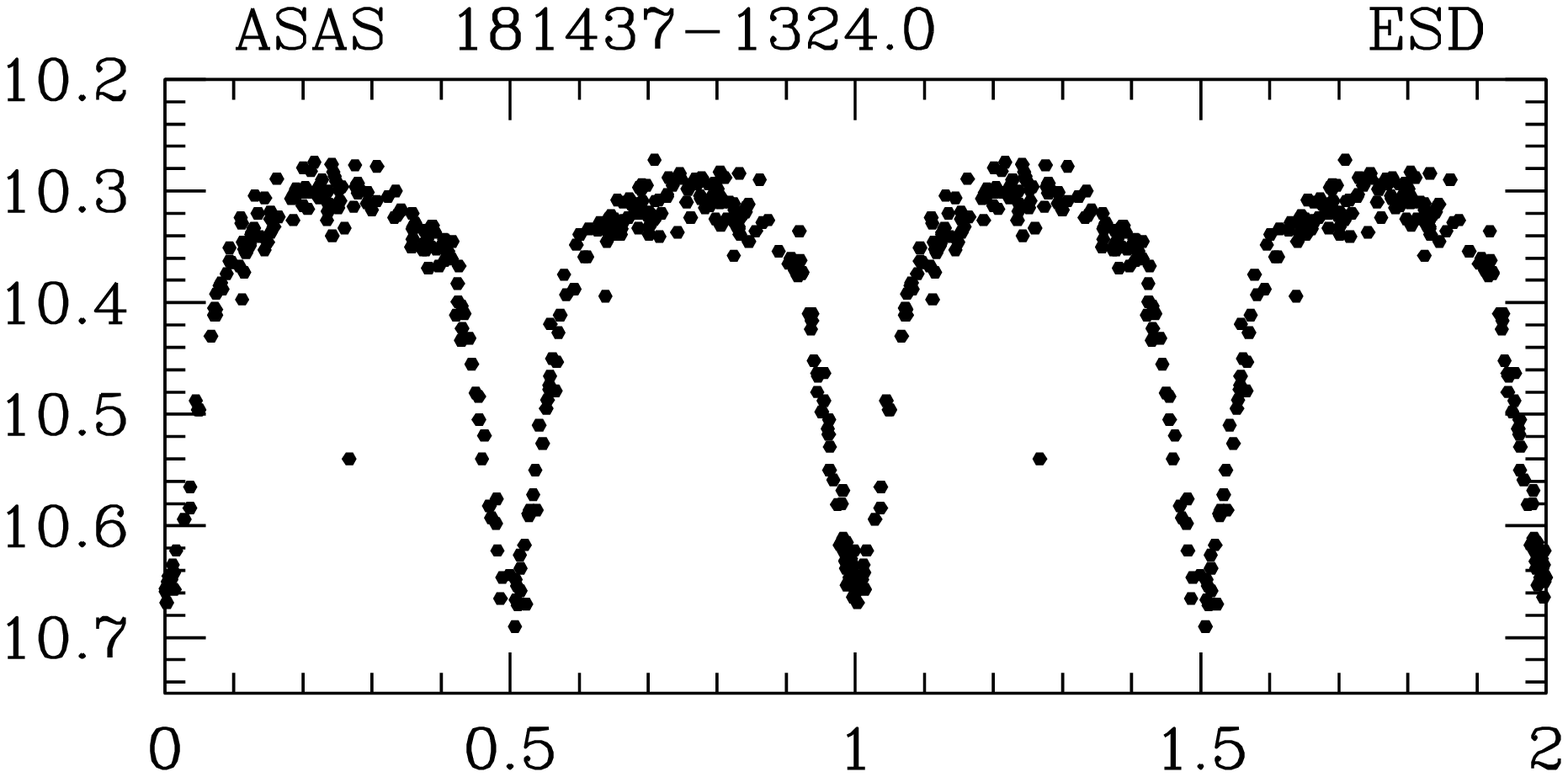}}&
  \resizebox{0.27\linewidth}{!}{\includegraphics*{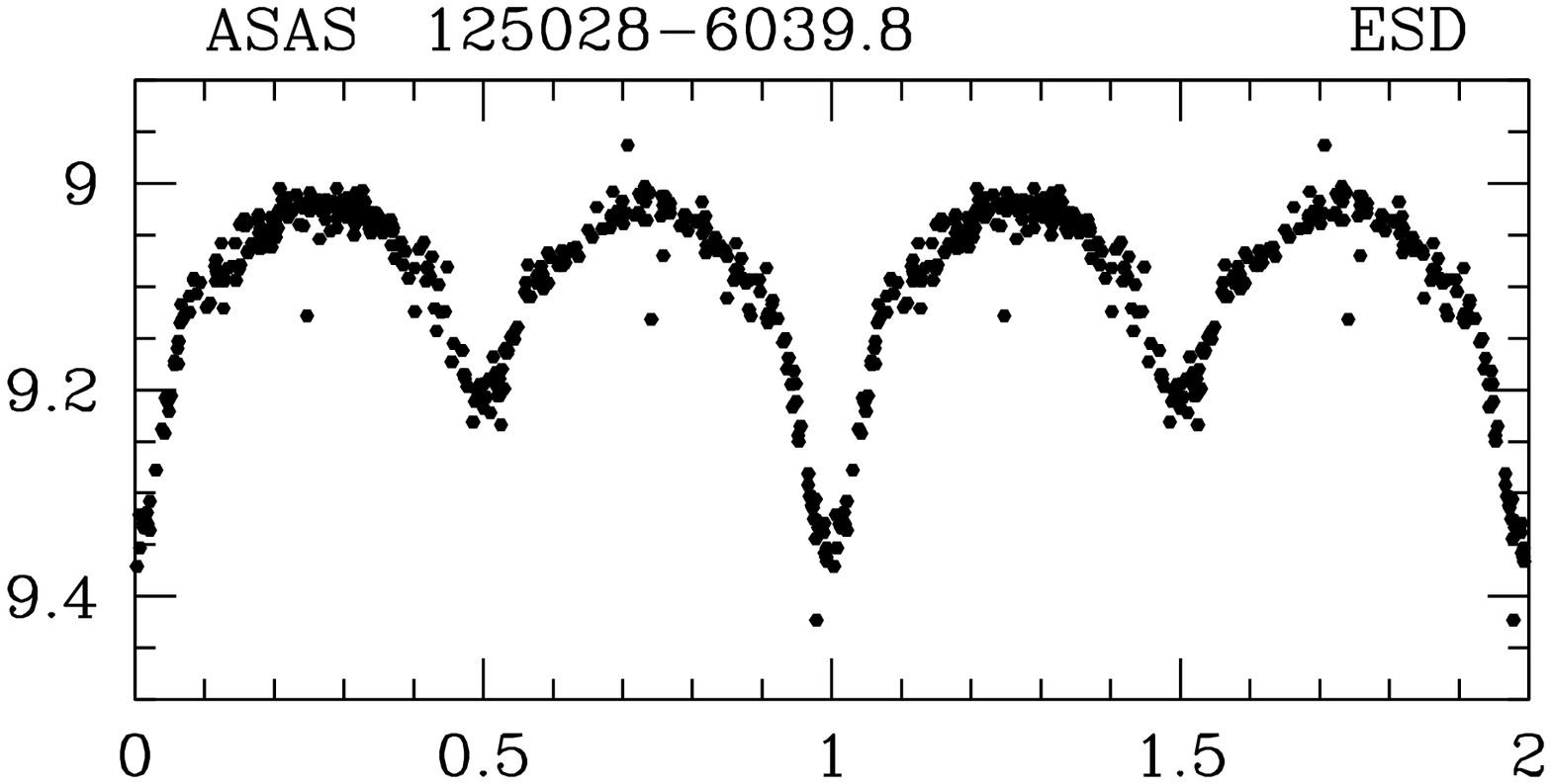}}&
  \resizebox{0.27\linewidth}{!}{\includegraphics*{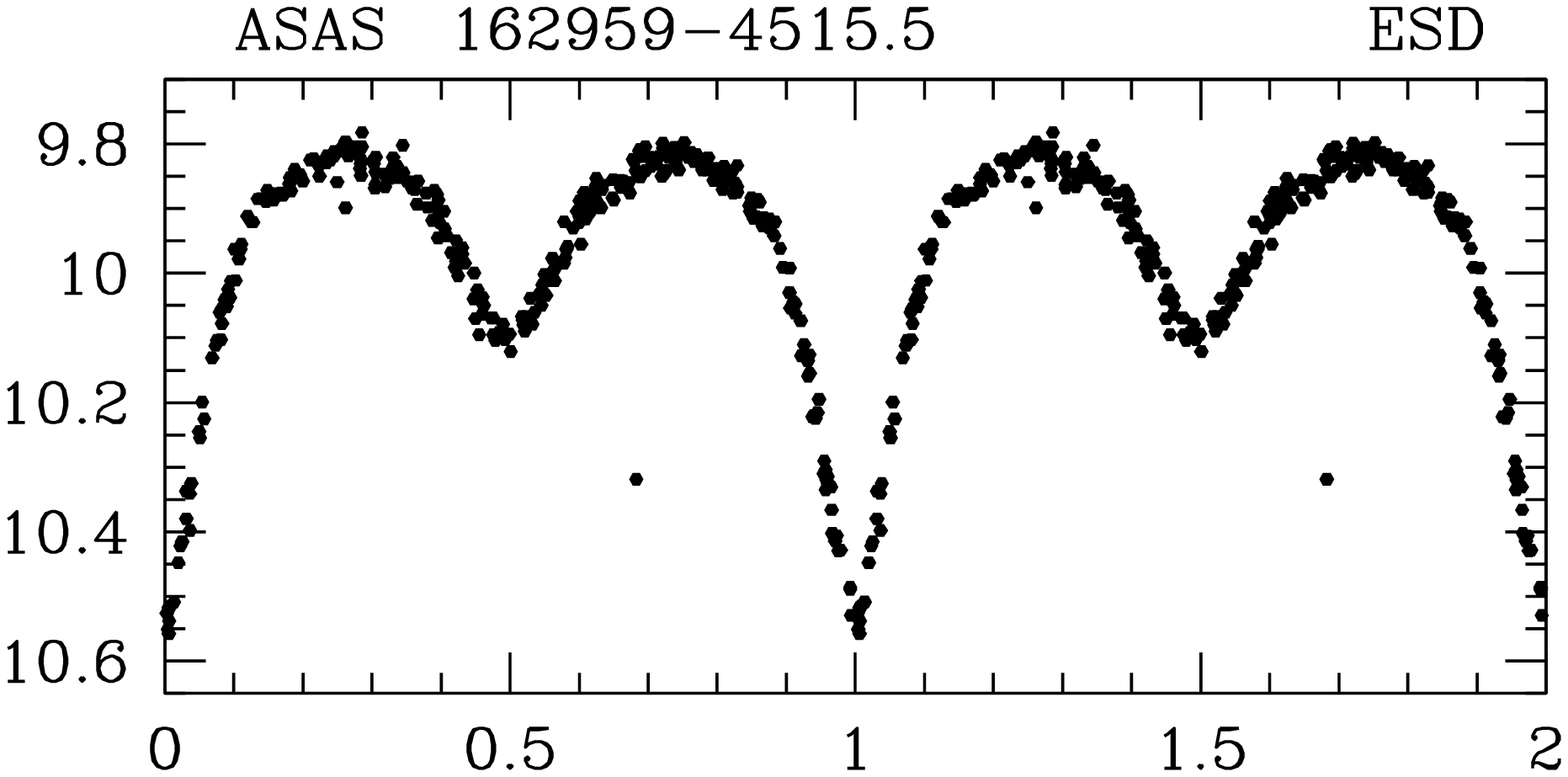}}\\
\vspace{-0.12\linewidth}
  \resizebox{0.27\linewidth}{!}{\includegraphics*{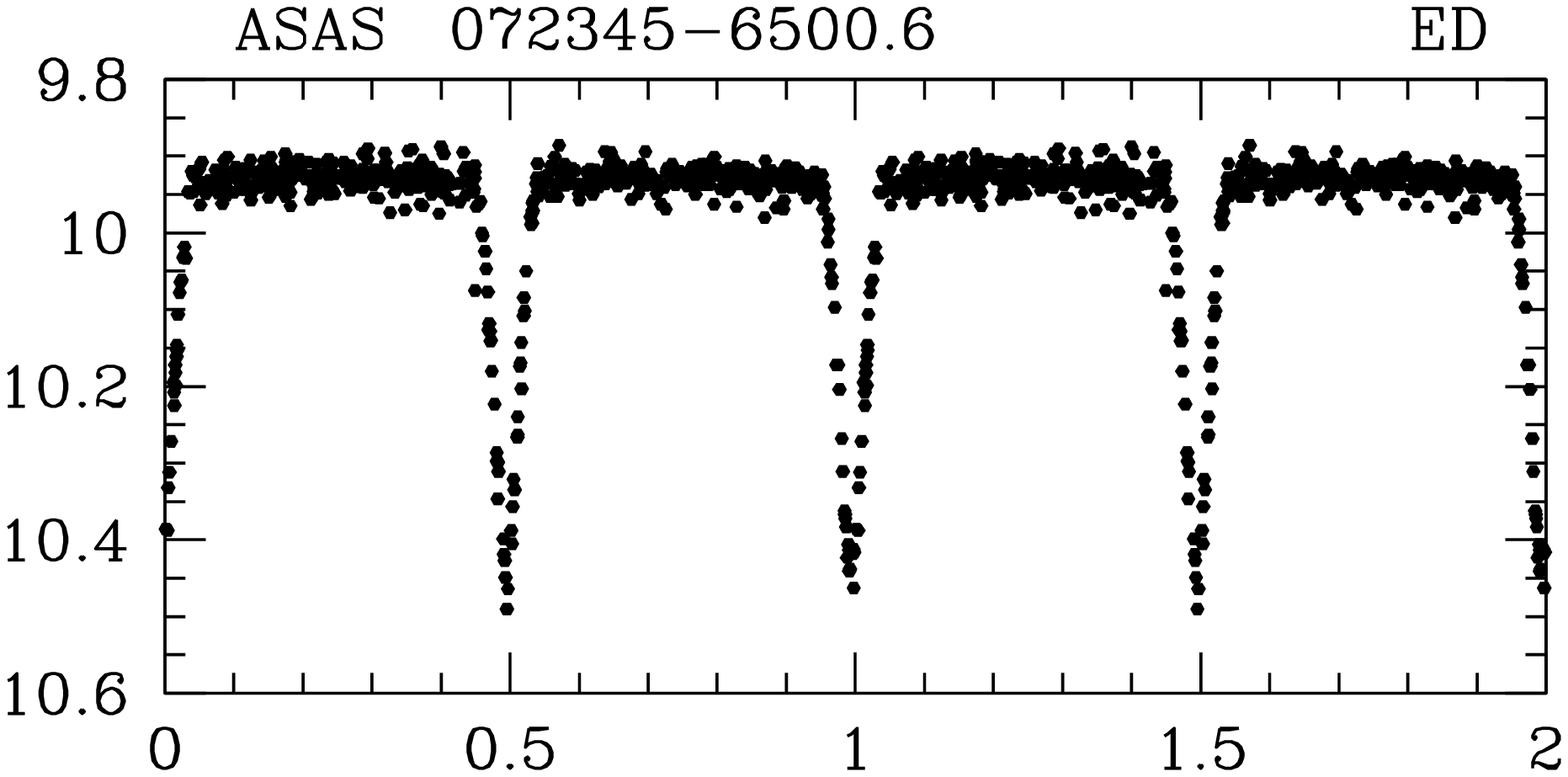}}&
  \resizebox{0.27\linewidth}{!}{\includegraphics*{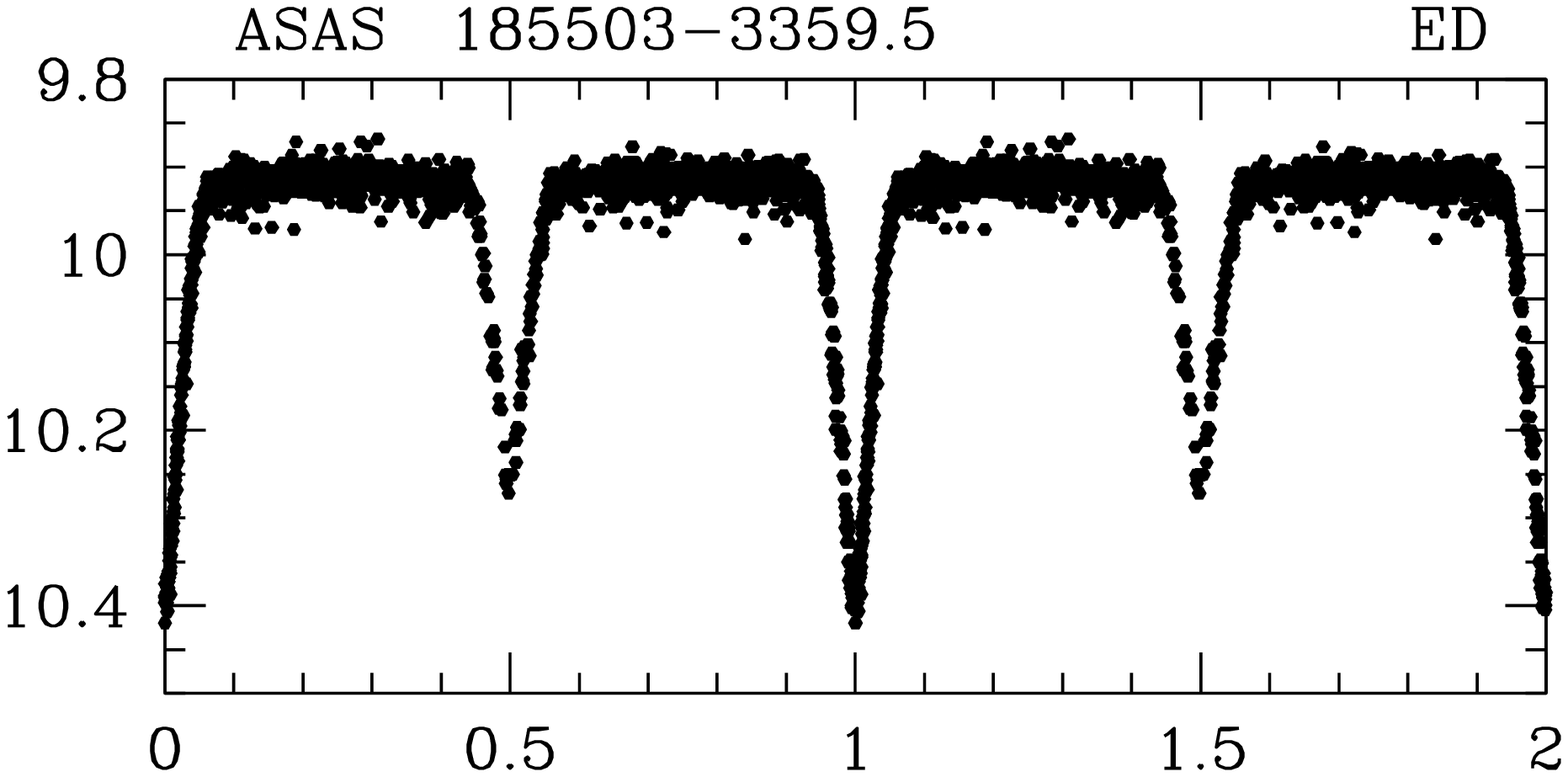}}&
  \resizebox{0.27\linewidth}{!}{\includegraphics*{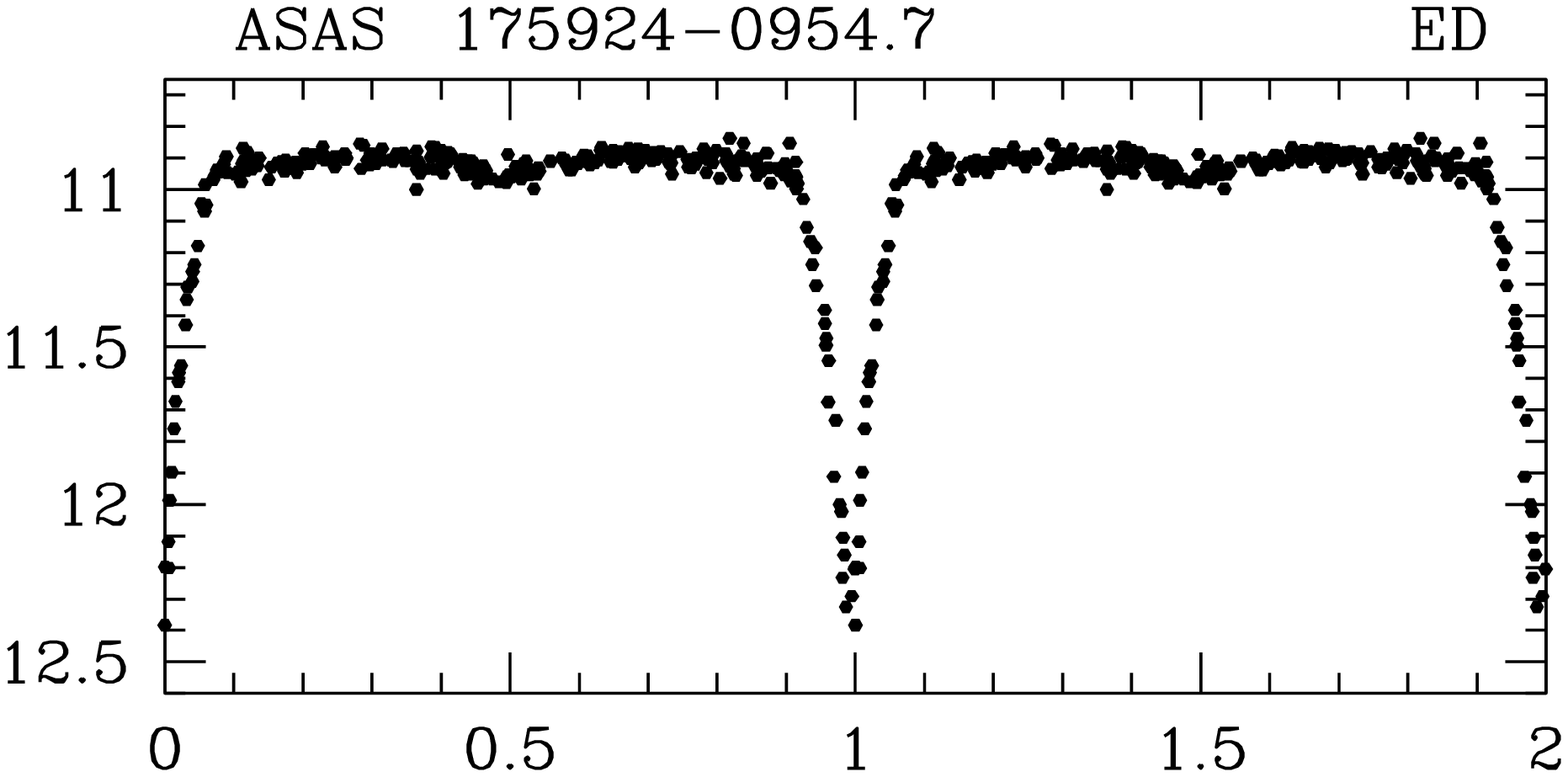}}\\
\vspace{-0.12\linewidth}
  \resizebox{0.27\linewidth}{!}{\includegraphics*{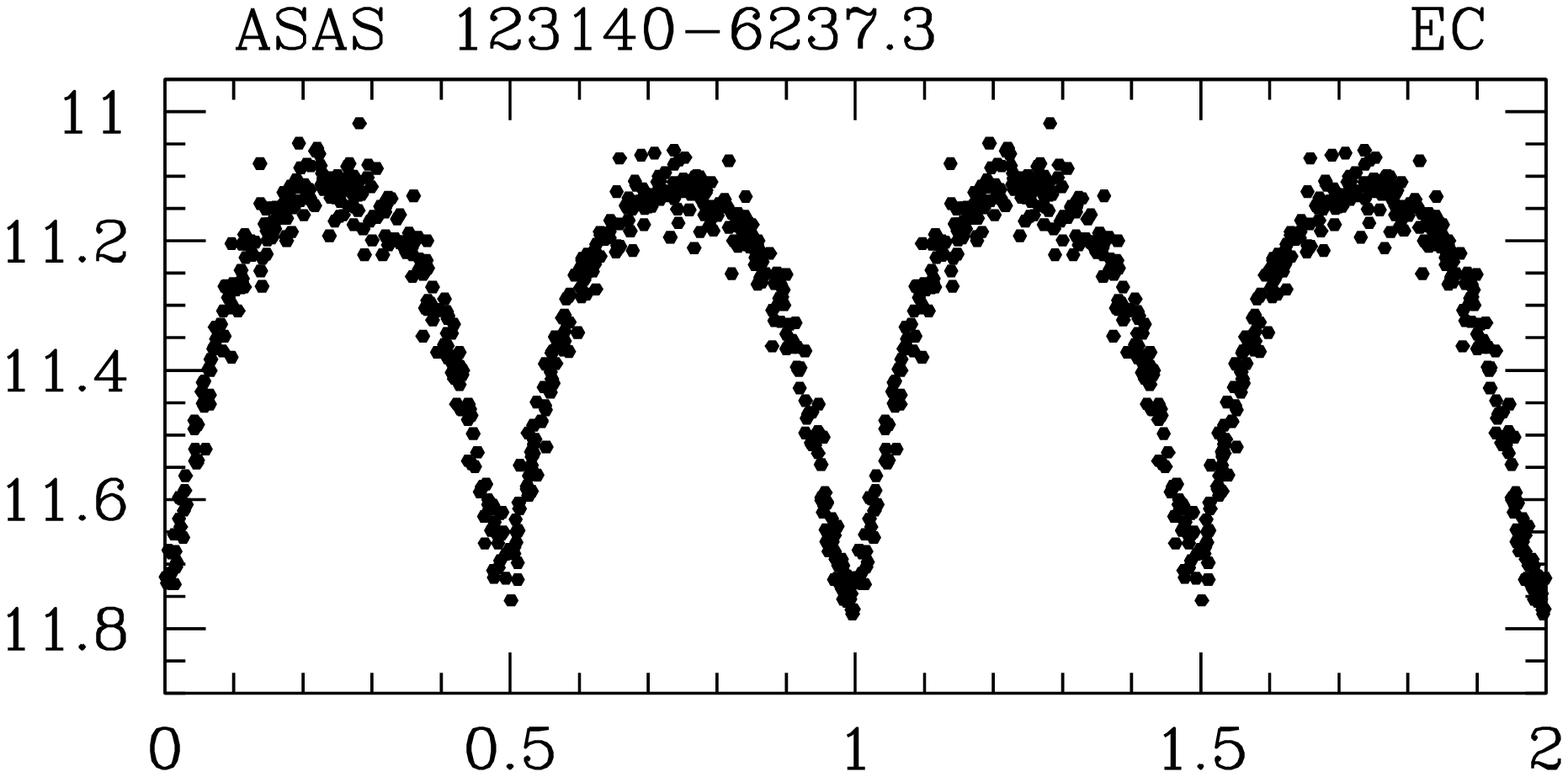}}&
  \resizebox{0.27\linewidth}{!}{\includegraphics*{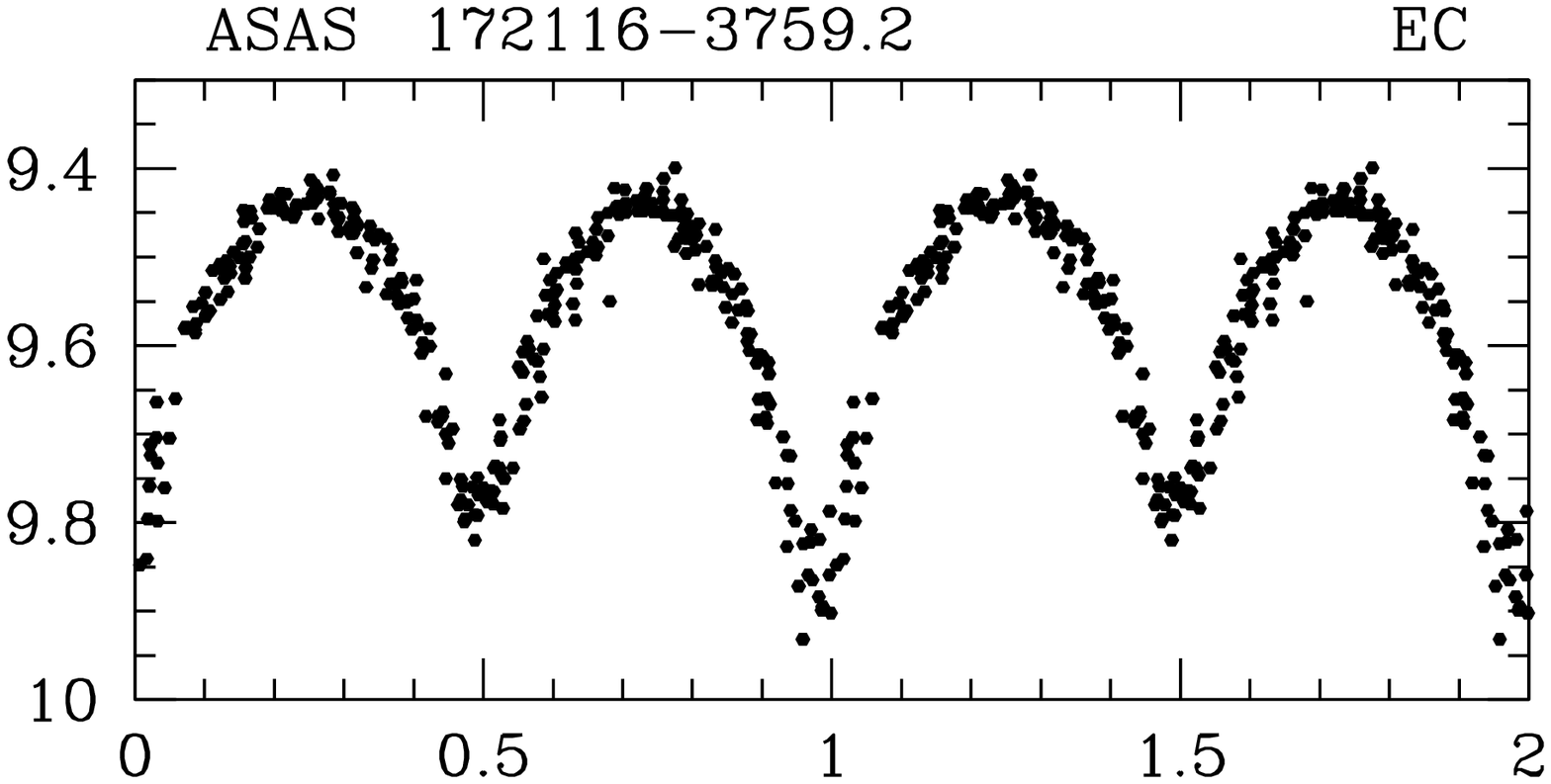}}&
  \resizebox{0.27\linewidth}{!}{\includegraphics*{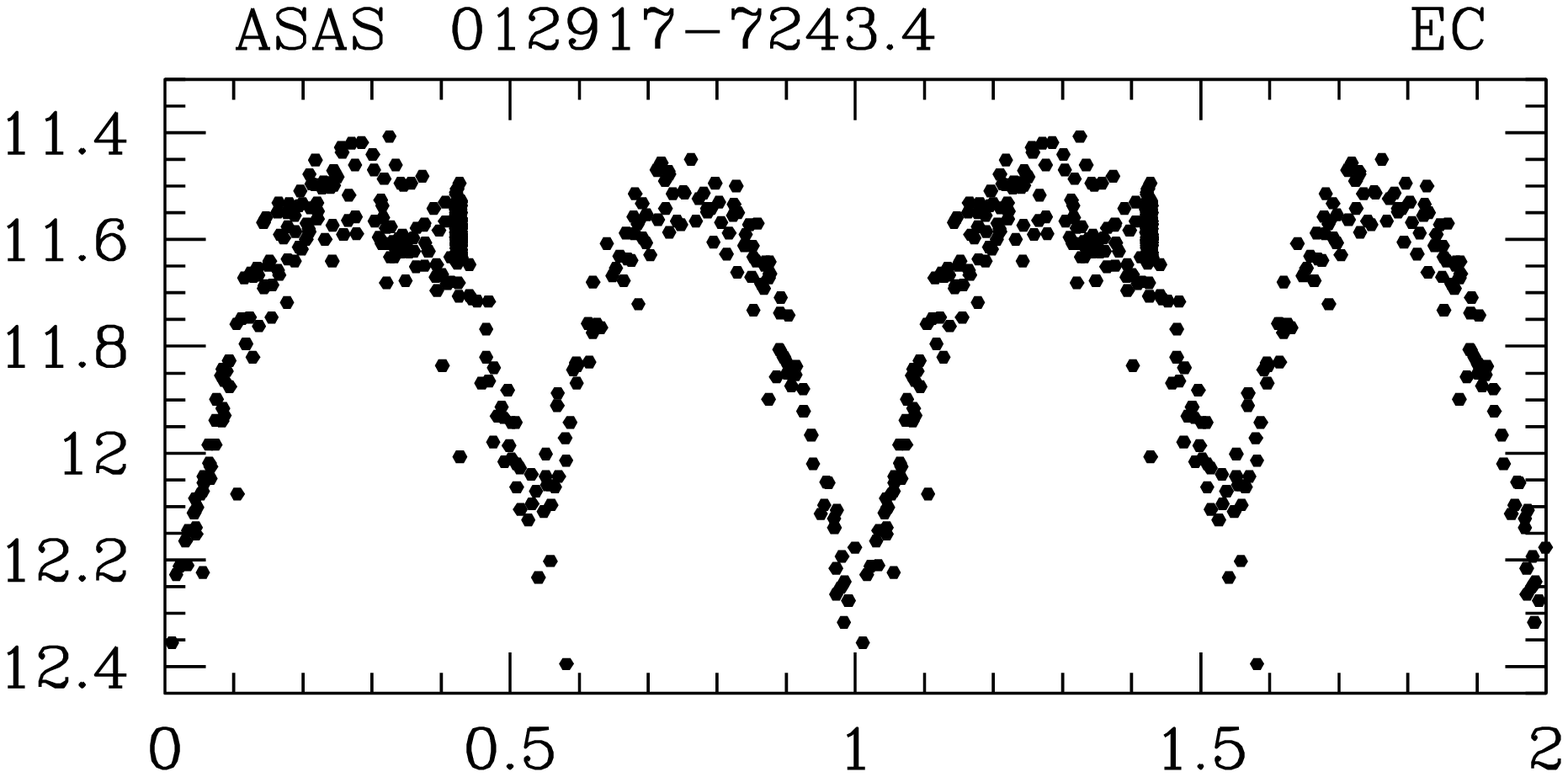}}\\
\vspace{-0.12\linewidth}
  \resizebox{0.27\linewidth}{!}{\includegraphics*{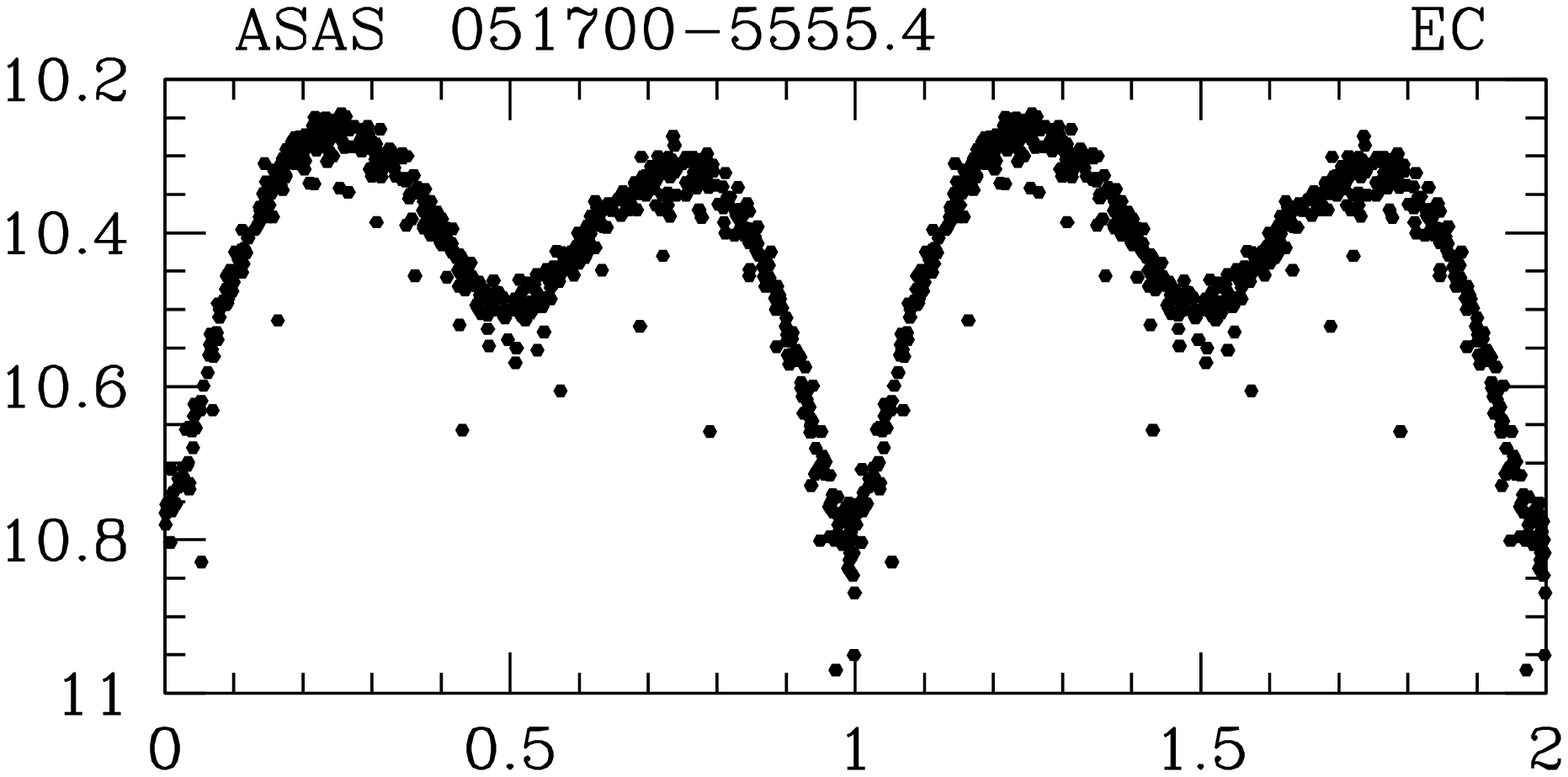}}&
  \resizebox{0.27\linewidth}{!}{\includegraphics*{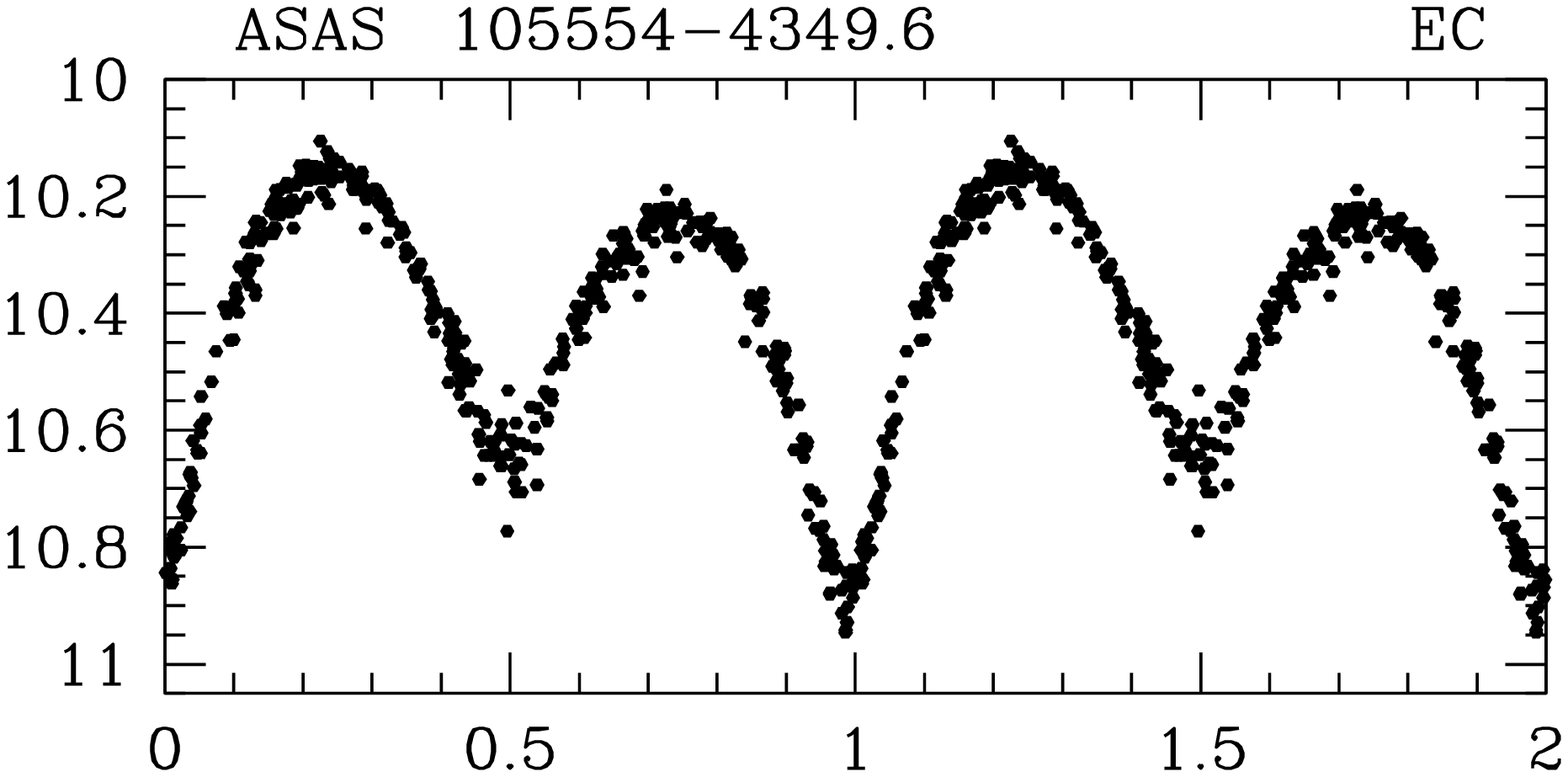}}&
  \resizebox{0.27\linewidth}{!}{\includegraphics*{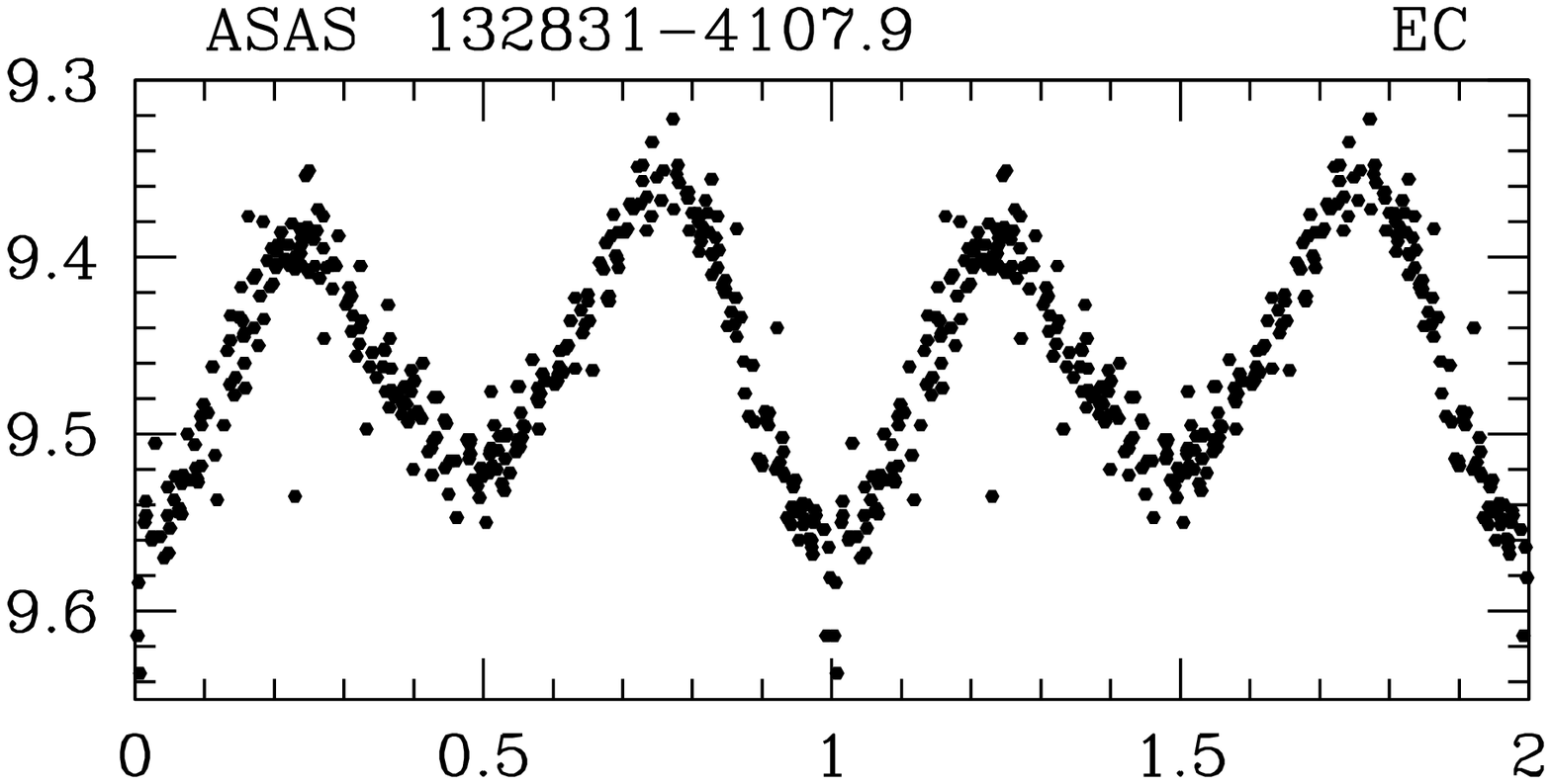}}\\
\vspace{0.08\linewidth}
\end{tabular}
\caption{
Examples of light curves of 9 contact (EC), 3 semi-detached (ESD)
and 3 detached binaries (ED) are shown.  All these are ASAS discoveries, even
though all of them are very bright.  Notice a distinct difference in
the light curve shapes between contact and a semi-detached binaries.
The first three rows are sorted by minimum depth difference, the fourth row
presents contact binaries with long end exceptionally long periods,
and the fifth with the distinct maximum height difference. See Table 1
for details.}
\end{figure*}

Some investigators (e.g. P. Eggleton) strongly prefer classification
of close binaries into EW, EB, and EA types, on the grounds that photometry
alone cannot provide unique geometry.  Our classification
into EC, ESD, and ED types should be considered preliminary, as we have
only single band photometry and no spectroscopic information for
thousands of our binaries.

\begin{table}
\begin{scriptsize}
\caption{The most important Fourier parameters for light curves from Fig.
5 are listed along with orbital periods. Each row is separated with a
horizontal line. The $a_{2}$ parameter can be easily
translated to the amplitude, $a_{1}$ and $a_{3}$ 
determine the difference between the two eclipses, and $b_{1}$
is related to the difference between the two maxima. 
Full Table 1 is given electronically as file {\tt Fourier.E} on our web page.
\label{tabelka}}
\begin{tabular}{c|c|c|c|c|c|c}  
ASAS ID & P$[days]$ & $a_{1}$ & $a_{2}$ & $a_{3}$ & $a_{4}$ & $b_{1}$  \\
\hline
125340-5010.6 & 0.4047 & -0.003 & -0.158 & -0.003 & -0.028 & -0.003 \\
111621-5427.2 & 0.4971 & -0.026 & -0.123 & -0.012 & -0.020 & -0.000 \\
220425-0603.4 & 0.7273 & -0.062 & -0.139 & -0.034 & -0.027 & 0.021 \\
\hline
181437-1324.0 & 1.3992 & -0.001 & -0.110 & -0.004 & -0.054 & 0.004 \\
125028-6039.8 & 2.0439 & -0.030 & -0.085 & -0.010 & -0.031 & -0.001 \\
162959-4515.5 & 1.1376 & -0.068 & -0.155 & -0.032 & -0.049 & 0.001 \\
\hline
072345-6500.6 & 3.1240 & -0.006 & -0.066 & -0.003 & -0.060 & -0.001 \\
185503-3359.5 & 1.1678 & -0.019 & -0.070 & -0.015 & -0.060 & 0.002 \\
175924-0954.7 & 2.3438 & -0.115 & -0.119 & -0.090 & -0.088 & 0.006 \\
\hline
123140-6237.3 & 1.9307 & -0.009 & -0.184 & -0.004 & -0.051 & 0.001 \\
172116-3759.2 & 5.2056 & -0.021 & -0.135 & -0.005 & -0.033 & 0.007 \\
012917-7243.4 & 180.60 & -0.050 & -0.190 &  0.016 & -0.052 & 0.041 \\
\hline
051700-5555.4 & 0.7904 & -0.059 & -0.126 & -0.028 & -0.021 & 0.020 \\
105554-4349.6 & 1.1541 & -0.034 & -0.185 & -0.019 & -0.030 & 0.037 \\
132831-4107.9 & 16.515 & -0.010 & -0.069 & -0.011 &  0.003 & -0.015 \\
\hline
\end{tabular}
\end{scriptsize}
\end{table}

A sample of the catalog of ASAS eclipsing binaries is shown in Table 2.
The full table has the information for all 11,099 binaries, and can be found
on our web page, in the file {\tt Vars.E}.

\begin{table*}
\begin{scriptsize}
\begin{minipage}{0.88\linewidth}
\caption{A sample of the catalog of ASAS eclipsing binaries is presented
here. Each star has its own ASAS designation (ID), equatorial
coordinates (RA and DEC), orbital period, epoch of minimum light (T0), and
ASAS classification. Additional information of other designation and
classification is also given, when available.  Full Table 2 is 
available electronically as the file {\tt Vars.E} in our web page.
\label{tabelka}}
\end{minipage}

\begin{tabular}{c|c|c|r@{.}l|c|c|c|c|c|c}  
ID & RA     & DEC    & \multicolumn{2}{|c|}{Period}   & T0       & V       & Amp     & Class & Other & Other \\
-  & (2000) & (2000) & \multicolumn{2}{|c|}{$[days]$} & 2450000+ & $[mag]$ & $[mag]$ &   -   & ID    & Class \\
\hline
065819+1028.4 & 06:58:19 & +10:28:24 & 0&447763 & 2387.63 & 11.04 & 0.28
& EC & & \\
070017+0202.2 & 07:00:17 & +02:02:12 & 1&397550 & 1873.07 & 10.53 & 0.57
& EC & V0460~Mon & EB/KE \\
211201$-$2003.2 & 21:12:01 & $-$20:03:12 & 0&352202 & 1872.84 & 13.22 & 0.34
& EC/ESD & & \\
025535$-$0219.9 & 02:55:35 & $-$02:19:54 & 0&79274 & 1920.00 & 11.59 & 0.54
& ESD & & \\
095930$-$6440.1 & 09:59:30 & $-$64:40:06 & 2&2428 &	1869.42 & 11.48 & 1.12 &
ESD & & \\
202110$-$4333.9 & 20:21:10 & $-$43:33:54 & 1&00218 & \ 1876.053 & 12.08 & 0.71
& ESD/ED & V2265~Sgr & EA \\
001855$-$7954.9 & 00:18:55 & $-$79:54:54 & 0&90310 & 1870.45 & 11.28 & 0.77
& ED & & \\
114500$-$7247.6 & 11:45:00 & $-$72:47:36 & 22&23037 & 1895.60 & 11.15 & 0.09
& ED & & \\
235052$-$2316.7 & 23:50:52 & $-$23:16:42 & 1&4023 &	1871.06 & \ 9.42 & 0.29 &
ED & & \\
\hline
\end{tabular}
\end{scriptsize}
\end{table*}

A subset of all ASAS results
related to binary stars was `frozen' for the epoch of this paper.
ASAS is an on-going project, with more data added every clear
night.  Therefore, as time goes on, the volume of data related to
eclipsing binaries will increase, and the quality will improve.
To make the results presented in this paper reproducible we decided to
provide `frozen' data.

Note: all light curves presented in Fig. 5 are for bright variables
discovered with ASAS survey.  They were not known before.
Just as expected, the majority of ASAS variables are new
discoveries (Paczy\'nski 1997).

\section{Statistics of contact binaries}
\label{sect:statistics}

All the statistics of this section is based on data provided in the file
{\tt Vars.E}, which can be found on our web page. Table 2 is a sample of this file.
In particular, the classification of ASAS binaries is given there.

The Galactic distribution of ASAS eclipsing binaries shows 
a significant difference, with detached binaries most strongly
concentrated to the galactic plane, while the short period contact
binaries have almost isotropic distribution.  It is well known
that there is a period - luminosity relation for W UMa stars
(Rucinski 1996, Klagyivik and Csizmadia 2004), and the Galactic 
distribution of those stars is not surprising.  The strong concentration 
of detached binaries to the Galactic plane implies these
are even more luminous and massive, on average.  

The binary period distribution
is shown in Fig. 6 for stars with the Galactic latitude 
${\rm |b| >  30^{\circ}}$ to reduce the luminosity bias.
The distribution of contact systems (EC) peaks near 0.37 days, it has 
a sharp cut-off at 0.2 days, and a long tail extending beyond 1 day.
In fact the tail extends to more than one hundred days, as it is
apparent in Fig. 8 and Fig. 10.  Also shown is the distribution of orbital
periods of semi - detached (ESD) and detached (ED) binaries.

\begin{figure}
\includegraphics[width=\linewidth]{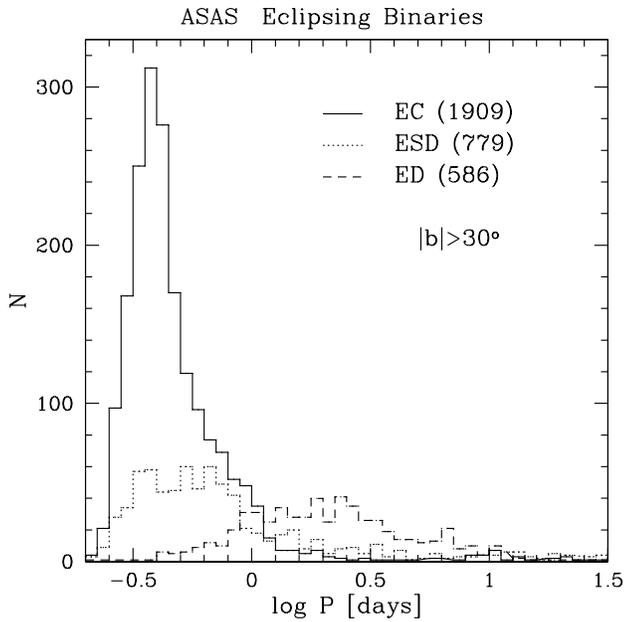}
\caption{
The distribution of periods of ASAS contact binaries (EC) at high Galactic 
latitude ${\rm |b| > 30^{\circ} }$ is plotted.
The distribution peaks near 0.37 days,
it has a sharp cut-off at 0.2 days, and a tail extending far beyond 1 day.
Also shown is the distribution of orbital
periods of semi - detached (ESD) and detached (ED) binaries.
Contact binaries outnumber other binaries for binary periods
shorter than 1 day.
}
\end{figure} 

The important topic ot this paper is the presentation of eclipse
depth of EC binaries, as this is directly related to the issue
of relaxation oscillations.  For contact binaries the 
depth of both eclipses is directly related to the Fourier
coefficients of the light curve.  The following
analytical formulae approximate these relations with better then $5\%$ 
accuracy for the majority of stars:
$$
{\rm D_{p} = -2 \left( 1.2a_{2} - 2a_{2}^2 + (a_{1}+a_{3}) \right) \hskip 1.0cm [mag] } ,
\eqno(1)
$$
$$
{\rm D_{s} = -2 \left( 1.2a_{2} - 2a_{2}^2 - (a_{1}+a_{3}) \right) \hskip 1.0cm [mag] } ,
\eqno(2)
$$
where ${\rm D_p }$ and ${\rm D_s }$ correspond to the depth of
the primary and secondary minimum, respectively.  Three examples
are presented in Fig. 7.

\begin{figure}
  \vspace{-3cm}
\begin{tabular}{c}
  \vspace{-3.3cm}
  \resizebox{0.8\linewidth}{!}{\includegraphics*{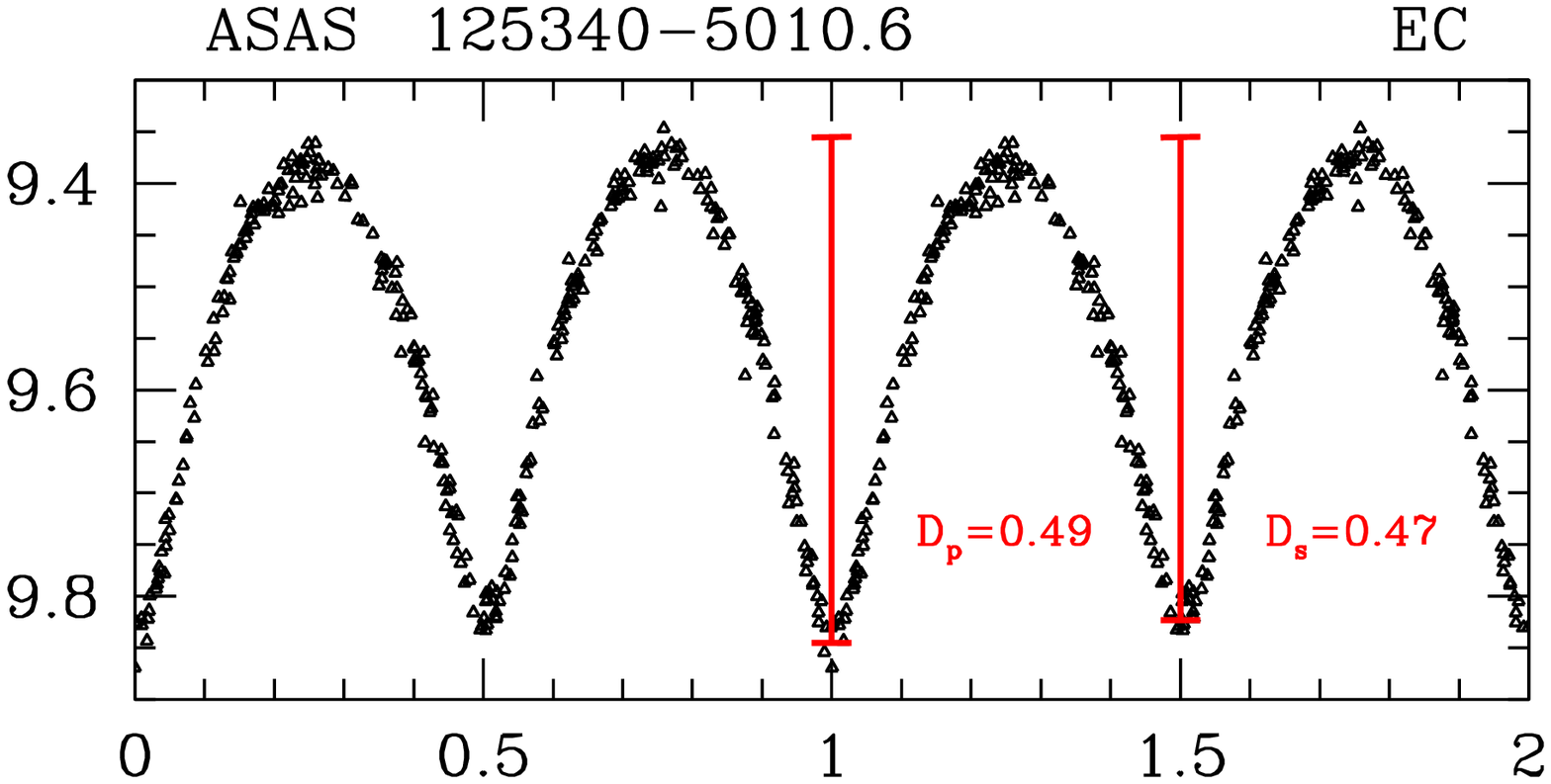}}\\
  \vspace{-3.3cm}
  \resizebox{0.8\linewidth}{!}{\includegraphics*{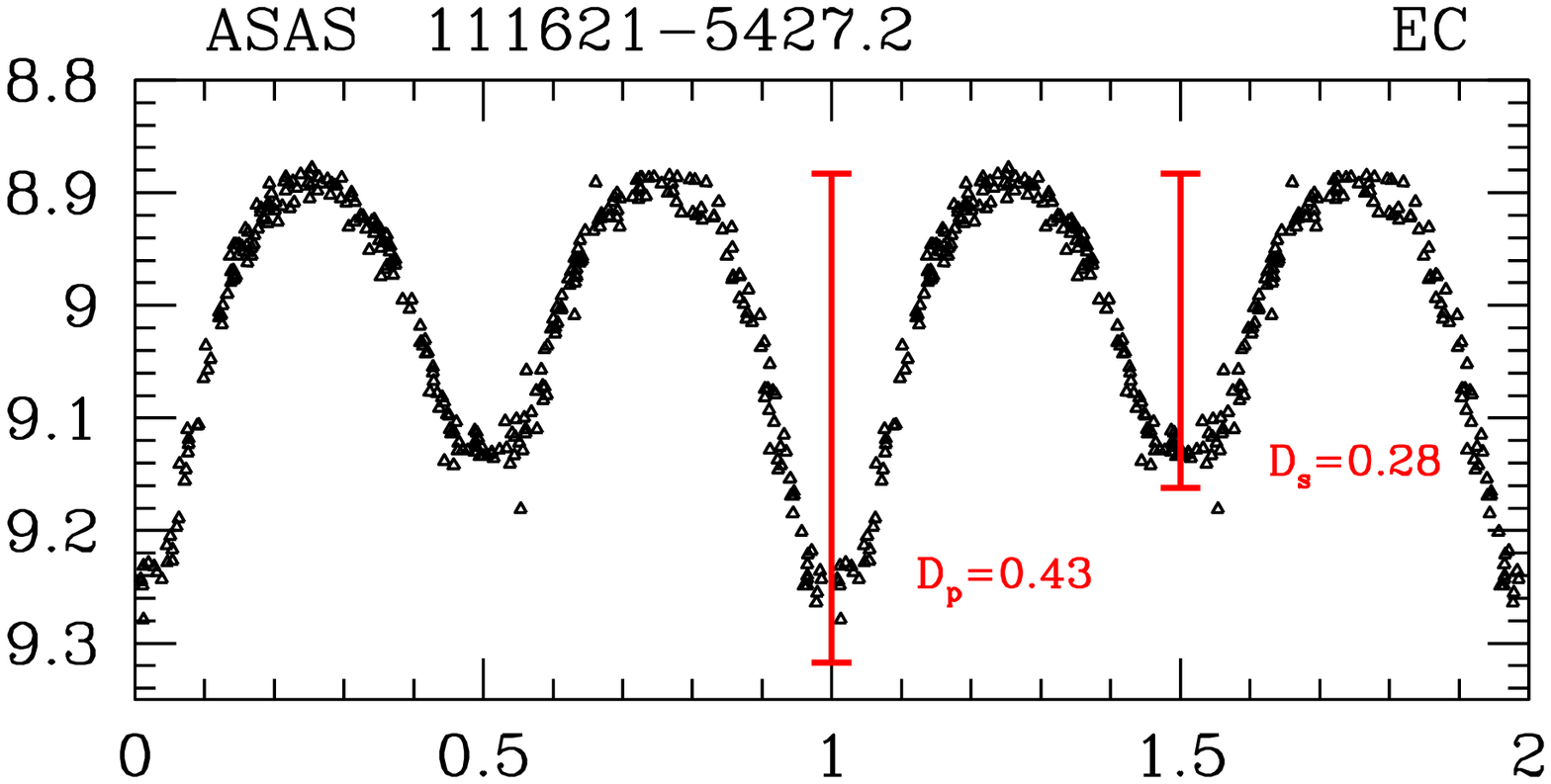}}\\
  \resizebox{0.8\linewidth}{!}{\includegraphics*{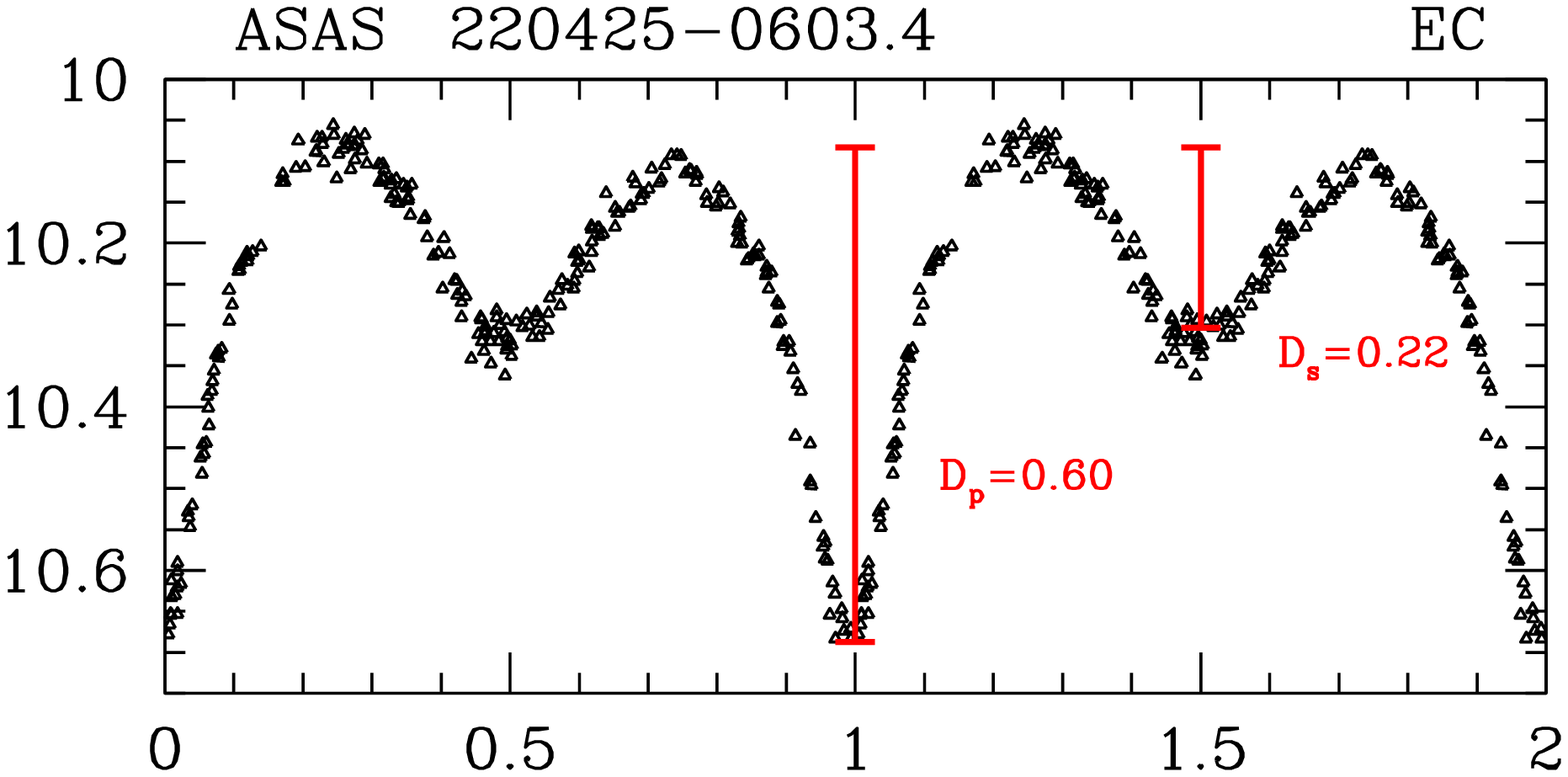}}\\
\end{tabular}
\caption{The first row of EC stars from Fig. 5 is shown. Vertical lines 
represent depth of a primary ($D_p$) and a secondary ($D_s$) eclipse
calculated with equations (1) and (2), respectively. Flat bottom eclipses
may result in a significant error, which is somewhat reduced 
in the $ {\rm {D_s} / {D_p} }$ ratio. }
\end{figure}

\begin{figure}
\includegraphics[width=\linewidth]{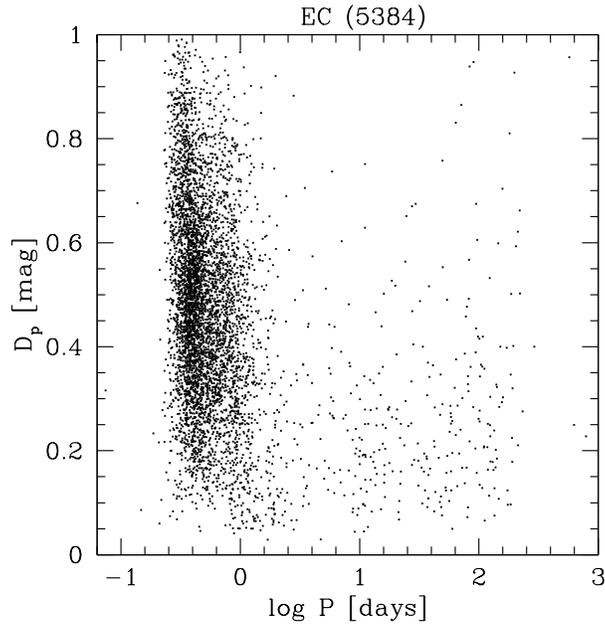}
\caption{
The distribution of primary eclipse depth as a function of orbital 
period for contact binaries.  ${\rm D_p }$ is the fraction of light
obscured in the primary eclipse.
}
\end{figure} 

The distribution of primary eclipse depth
with binary period is shown in Fig. 8.
A histogram of primary eclipse depth integrated over binary period
is shown in Fig. 9.  Note that while the vast majority of W UMa
binaries have periods in the range ${\rm 0.2 < P < 1.2 }$ days,
there are contact binaries with periods over 100 days, as it is
apparent in Fig. 5 and Fig. 8.

\begin{figure}
\includegraphics[width=\linewidth]{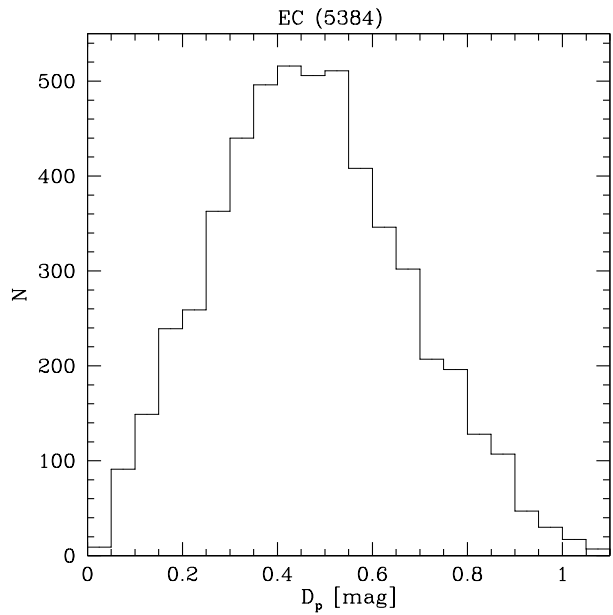}
\caption{
The distribution of primary eclipse depth integrated over all periods.
}
\end{figure} 

The eclipse depth ratio is shown in Fig. 10 for EC binaries, and in Fig. 11
for EC and ESD binaries.  These are important figures.  They demonstrate
that some EC binaries are in good thermal contact, as 
the depth of secondary eclipse is almost the same as the primary.
But many other EC binaries have the two eclipses which are very unequal,
implying that they are not in thermal contact, yet their geometry
is close to that expected for a contact system.  Finally, there
are ESD binaries, in which the geometry is very different, and
of course the two eclipses have usually different depth.
Qualitatively, this is
just what was expected in the models with relaxation oscillations
(cf. Lucy 1976, Flannery 1976, Robertson and Eggleton 1977, Yakut
and Eggleton 2005).  There is no shortage of stars without
thermal contact.  Combining ASAS data with models of relaxation
oscillations should provide quantitative verification of the theory,
but this task is beyond the scope of our paper.

\begin{figure}
\includegraphics[width=\linewidth]{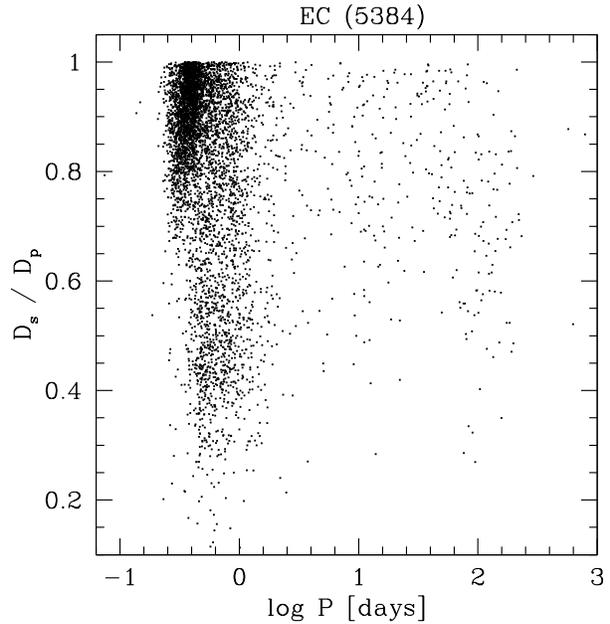}
\caption{
The ratio of eclipse depth is shown for contact binaries as a function
of their period.  Stars in thermal contact have the eclipse
depth ratio close to one.
}
\end{figure}

\begin{figure}
\includegraphics[width=\linewidth]{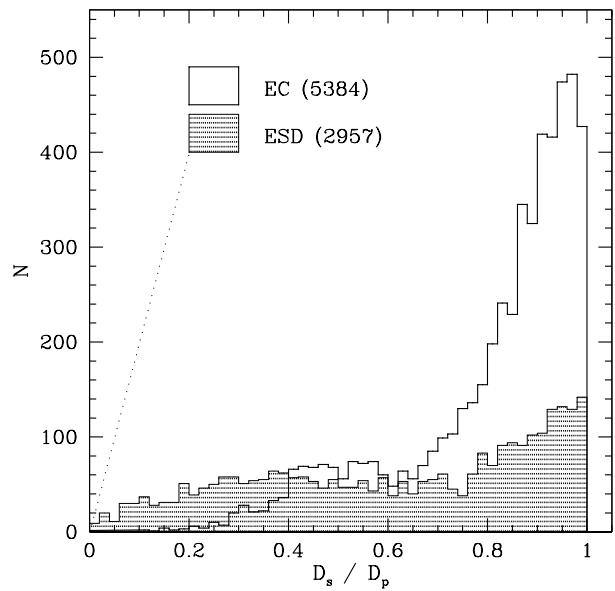}
\caption{
The ratio of the eclipse depths, integrated over binary periods, 
is shown for contact (EC) and semi-detached (ESD) systems.
Stars in a thermal contact have the eclipse
depth ratio close to one.  Note the step in the histogram
at ${\rm D_s/D_p \approx 0.7 }$  The same feature is also
apparent in Fig. 10 at ${\rm \log P \approx -0.4 }$.
}
\end{figure}

Contact binaries often have maxima of different
height, with the maximum following the primary (i.e. deeper)
eclipse being either higher or lower than the maximum following
the secondary eclipse.  The distribution of the difference
is shown in Fig. 12. 
The positive value of the ${\rm b1/|a2| } $
parameter indicates that the maximum following the primary eclipse is 
brighter of the two.  The asymmetry in distribution, i.e.
the excess of positive values of ${\rm b1/|a2| }$, 
is known as the O'Connell (1951)
effect (cf. also Ruci\'nski 1997b): the maximum following
the primary eclipse is on average brighter, presumably due
to stream of gas flowing between the two stars.  An extreme
case of this phenomenon is shown in V361 Lyr (Kaluzny 1991).

\begin{figure}
\includegraphics[width=\linewidth]{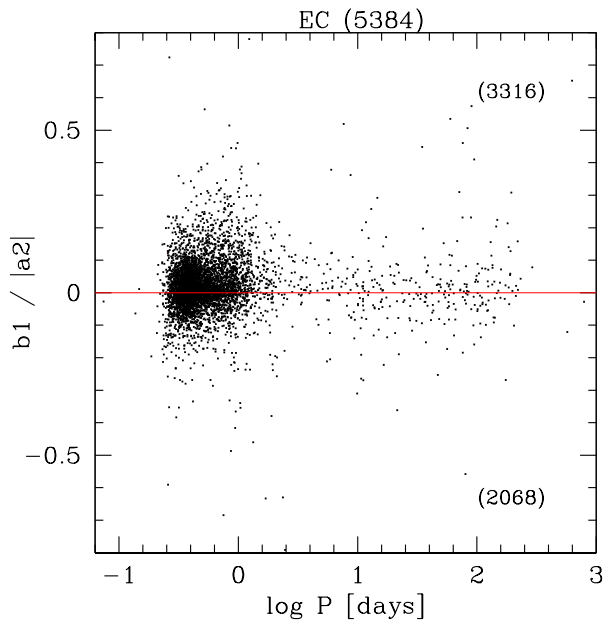}
\caption{
The distribution of light curve asymmetry of contact binaries is 
shown - this is known as O'Connell (1951) effect.  
Most stars have their light maxima of approximately equal height, but 
there is an asymmetry in the distribution: the maxima that follow
the primary minimum are on average higher than the maxima
preceding the primary eclipse.  The O'Connell effect is
likely a consequence of gas streams in these binaries. 
}
\end{figure}

\section{Discussion}
\label{sect:discuss}

Our conclusion, based on the distribution of eclipse depths
(Fig. 10, 11), is that 
relaxation oscillations, first proposed by Lucy and
Flannery, are real.  There is no shortage
of binaries corresponding to no thermal contact,
with a very different depth of the two eclipses.  Model calculations of
the type recently done by Yakut and Eggleton (2005), when
combined with ASAS data, should allow a quantitative verification
of the theory.

A very large number of contact or near contact systems 
reveal features never noticed before: 
the distribution of eclipse depth ratios as shown in Fig. 10
and Fig. 11 has a distinct break around ${\rm log P \approx -0.4 }$
and ${\rm D_s/D_p \approx 0.7 }$.  We do not speculate on the
origin of this feature, but we bring this break
to the attention of our readers.  This break is
best seen in systems with the most robust classification.

The most surprising result of this paper is presented in Fig. 6.
A traditional view for the origin of W UMa contact binaries
is to assume that they come from detached binaries of comparable
periods.  For the first time we have approximately complete
statistics of binaries of all types with orbital periods
shorter than one day, and there are very few detached binaries.
Obviously, they are more difficult to find than 
either contact or semi - detached systems.
As times goes on the statistics of ASAS detached binaries
will improve, and a statistical analysis will tell us if
there is a problem with the origin of W UMa stars.  
At this time the contact systems seem to appear "out of nowhere".

In about a year we shall have I-band data for our eclipsing binaries.
This will make it possible to be more quantitative about the
distribution properties, including the distribution in our Galaxy.
Also, it will be much easier to quantify the impression about
a shortage of detached binaries and the space density of systems
of different types: EC, ESD, and ED.  At this time it is premature for us 
to speculate about the outcome of binary statistics while we wait
for the I-band data. 

While this is an observational paper, written to promote the usefulness
of the ASAS catalog of variable stars, we are tempted to speculate about
possible interpretation of Fig. 6, which was so surprising to us.
We are not consistent with the previous paragraph, but the temptation
is hard to resist.  The following is a speculative hypothesis.

There has been a gradual emergence of the notion that contact
systems have companions (Ruci\'nski and Kaluzny 1982, Chambliss 1992,
Hendry and Mochnacki 1998, also Tokovinin 2004).
Recently. Pribulla and Ruci\'nski (2005) found that up to 50\% of W UMa
binaries have companions.
This opens up a possibility that Kozai (1962) cycle operates
in some such triples, as suggested by Kiseleva, Eggleton,
and Mikkola (1998), and by Eggleton and Kiseleva-Eggleton (2001).
The mechanical three-body orbital evolution,
with a large range of inclinations of the two orbits, inner and
outer, may induce large variations in the eccentricity of
the inner binary.  In time the eccentricity may become large
enough to make a contact system out of the inner binary.

In this scenario the inner binary has the initial period that is 
relatively long, and in most cases no eclipses would be detectable.
During a Kozai cycle the inner binary occasionally has a chance
to reach a physical contact: it may either merge forming
a single star, or it may become a W UMa star.  This process
is somewhat similar to the model of formation of close
binaries in globular clusters (Pooley et al. 2003, and references therein).

The surprisingly small number of short period detached
eclipsing binaries as seen in Fig. 6 may indicate that
the Kozai cycle is not just a curiosity, but it may be
an important channel for forming W UMa stars.
This possibility cannot
be rejected without careful analysis.  After all W UMa
stars are rare, with the local space density of just 0.2\%
of the main sequence stars (Rucinski 2002, and references therein).

\section*{Acknowledgements}

We are very grateful to P. P. Eggleton, D. Fabrycky, S. Ruci\'nski and 
K. Stepie\'n for many helpful discussions.  

The ASAS project was supported to a large extent by the generous
donation of William Golden.
This work was partially supported by the Polish MNiI  grant
no. 1P03D-008-27.

\bsp
\label{lastpage}

\end{document}